\documentclass{article}
\usepackage[utf8]{inputenc}
\pdfoutput=1
\usepackage{amsmath}
\usepackage{bm}
\usepackage{standalone}
\usepackage{import}
\usepackage{tabularx}
\usepackage{graphicx}
\usepackage{lscape}
\usepackage{longtable}
\usepackage{pgfplots}
\usepackage{tikzscale}
\usepackage{hyperref}

\usetikzlibrary{arrows, automata}

\usepackage{setspace}
\usepackage{array,multirow}
\usepackage{authblk}
\usepackage{import}

\usepackage{tcolorbox}

\usepackage{natbib}

\newcolumntype{P}[1]{>{\raggedright\arraybackslash}p{#1}}

\defcitealias{igme2020}{IGME, 2020}
\defcitealias{wang2017global}{Wang et al., 2017}
\defcitealias{alkema_bayesian_2017}{Alkema et al., 2017}
\defcitealias{mmeig2015}{UN MMEIG, 2015}
\defcitealias{mmeig2019}{UN MMEIG, 2019}
\defcitealias{alkema_global_2014}{Alkema et al., 2014}
\defcitealias{you_global_2015}{You et al., 2015}
\defcitealias{dicker_global_2018}{Dicker et al., 2018}
\defcitealias{li_changes_2019}{Li et al., 2020}
 
\setcitestyle{authoryear, open={((},close={)}}

\title{Temporal models for demographic and global health outcomes in multiple populations: Introducing a new framework to review and standardize documentation of model assumptions and facilitate model comparison }

\author[1]{Herbert Susmann}
\author[2]{Monica Alexander}
\author[1]{Leontine Alkema\thanks{The work was supported by the Bill \& Melinda Gates Foundation. Contact: Herbert (Herb) Susmann (hsusmann@umass.edu). This work is licensed under CC-BY 4.0. }}

\affil[1]{Department of Biostatistics and Epidemiology \protect\\ University of Massachusetts Amherst \protect\\ 715 North Pleasant Street \protect\\ Amherst, MA 01003}
\affil[2]{Departments of Statistical Sciences and Sociology \protect\\ University of Toronto \protect\\ 700 University Avenue \protect\\ Toronto, ON M5G 1Z5}
\date{\today}

\begin{document}

\maketitle
\onehalfspacing

\section*{Abstract}

There is growing interest in producing estimates of demographic and global health indicators in populations with limited data. Statistical models are needed to combine data from multiple data sources into estimates and projections with uncertainty. Diverse modeling approaches have been applied to this problem, making comparisons between models difficult. We propose a model class, Temporal Models for Multiple Populations (TMMPs), to facilitate documentation of model assumptions in a standardized way and comparison across models. The class distinguishes between latent trends and the observed data, which may be noisy or exhibit systematic biases. We provide general formulations of the process model, which describes the latent trend of the indicator of interest. We show how existing models for a variety of indicators can be written as TMMPs and how the TMMP-based description can be used to compare and contrast model assumptions. We end with a discussion of outstanding questions and future directions.

\bigskip


\newpage

\newpage
\section{Introduction}

Population-level measures of demographic and health indicators over time are essential to identify which populations are the most disadvantaged, where progress has been made or is stalling, and help to inform resource allocation. Important indicators include measures of health and mortality at different ages, fertility, family planning measures and migration. Assessing changes in quantities over time is useful for both cross-country and within-country comparisons to see how outcomes have changed in the past and how they are likely to change in future. 

In practice, producing estimates and projections of demographic and health indicators may not be straightforward due to the lack of good-quality data. Well-functioning civil registration and vital statistics systems (CRVS), which are usually the most reliable source of demographic information, do not exist in most countries worldwide, making reliable data on births, deaths, population and other health outcomes difficult to obtain. This is particularly the case in low-income countries, where the burden of mortality is often the highest. In many cases, there are missing observations over time, and in some populations we may have no observations. Data that do exist are sometimes of very poor quality, and there may be issues reconciling different observations of the same outcome. For example, in the absence of CRVS data, large-scale national surveys such as the Demographic and Health Surveys and Multiple Indicator Cluster Survey  \citep{croft2018guide, khan2019multiple} are common sources of information about health and mortality; however, even if the surveys cover an overlapping period, measurements derived from the two sources are not necessarily in agreement. Figure \ref{fig:gbd_b3_comparison} illustrates this problem by showing the data used for estimating under-five mortality rates in Senegal, which come from a variety of data sources and are often not in agreement with one another.

As such, statistical methods are usually required to model indicators over time because the data we observe are incomplete or paint an imperfect picture of what we are actually trying to estimate. Statistical models for the temporal estimation of demographic and health indicators have been used for decades, with some of the earliest efforts producing stochastic forecasts of mortality and fertility in the US \citep{lee1992modeling, lee1994stochastic}. As computational speed and power has increased, so too has the complexity of models. In the global health context, the era of monitoring progress towards Development Goals has elicited the creation of statistical models to capture non-linear trends in a wide range of indicators across a variety of data-sparse situations; see for example work in estimating under-5 mortality rates \citep{alkema_global_2014, you_global_2015, dicker_global_2018, li_changes_2019} and maternal mortality \citep{alkema_global_2016, alkema_bayesian_2017, kassebaum2016global}. In addition, as the focus of studying population-level outcomes has shifted to understanding inequalities across key sub-groups, the methods used to monitor trends at the macro level have also been applied to more finer-grained geographies, particularly where issues of small populations become apparent \citep{alexander_flexible_2017,  burstein2019mapping, li_changes_2019}. Statistical methods to account for and smooth temporal variation used are wide ranging, and include ARIMA models \citep{alkema_bayesian_2017}, penalized splines regression \citep{alexander_global_2018}, and Gaussian process regression \citep{dicker_global_2018}. Models may be hierarchical, spatial, and include explanatory covariates. 

On the outset, the statistical techniques used to model important demographic and health indicators appear to be diverse and highly dependent on the specific outcome being estimated, the background of the researchers, and the intended audience and users of the estimates. However, when viewed in the broader context of common modeling goals --- to produce reliable estimates and projections over time, accounting for missing data and multiple data sources, and to give an idea of the level of uncertainty in estimates and projections --- a large set of models can be explained in reference to a general model class. 

In this paper, we propose such a model class, ``Temporal Models for Multiple Populations" (TMMPs), that encompasses many existing demographic and health models. The model class has two main components: a process model and a data model. The aim of the process model is to capture trends over time in the underlying outcome of interest, which is itself unobserved. The process model is composed of a systematic and covariate parts, which may capture the relationship between the outcome and covariates, and a temporal smoothing component, which allows for data-driven trends over time. The data model relates the observed data to the underlying latent outcome of interest, taking into account different types of error that may be present in different data sources. Considering models in this class makes clear the distinction between the goal of capturing systematic, underlying trends in the `true' outcome of interest, and the goal of accounting and adjusting for different types of measurement error in observations.

After introducing statistical notation for the TMMP model class, we describe several existing models of a range of health and demographic indicators to illustrate how they fit into our proposed model class, and how these standardized descriptions can be used for comparison across models. The first two models described both aim to produce under-five country-level estimates of child mortality, but employ markedly different strategies \citep{alkema_global_2014, dicker_global_2018}. Other models discussed are: a model of contraceptive use rates, which provides an example of a model that captures a transition process \citep{cahill_modern_2018}; a model to estimate neonatal mortality rates globally, which includes a strong association with an underlying covariate \citep{alexander_global_2018}; estimating maternal mortality rates, which combines a multilevel mixed model and an ARIMA time series model \citep{alkema_bayesian_2017}; and a model of sub-national age-specific mortality, which shows how a more traditional demographic modeling approach can be expressed in our model class  \citep{alexander_flexible_2017}. 

The rest of this paper is structured as follows. Section~\ref{section:problem} formalizes the setting and describes the core statistical task at hand. Section \ref{section:case_study} describes two existing models for the same indicator, illustrating how their original notation makes it difficult to directly compare their modeling assumptions and motivating the use of a unifying modelling class. Section \ref{section:framework_overview} presents an overview of the proposed framework for TMMPs, introducing the distinction between process and data models, before examining the process model in more detail and reviewing strategies for parameter estimation, including hierarchical modeling. Section~\ref{section:case_study_revisited} returns to the models of Section \ref{section:case_study}, casting them in our model class to show how it facilitates their comparison. Additional detailed examples are included in Section~\ref{section:examples}. We discuss open problems and future directions for research in Section  \ref{section:discussion}.

\section{Problem statement}
\label{section:problem}
In this section we describe the overarching statistical problem that the models in our proposed class intend to solve. They are designed to produce estimates and future projections of a particular outcome for a set of populations over a certain time period. For example, the modellers may be interested in estimating and projecting child mortality for every UN-member country
over the period 1990 until the current year. 

In general, the outcome of interest could be any population-level demographic or health indicator. The populations of interest could include national, regional or subnational populations, or population subgroups based on some other characteristic (e.g. sex, wealth quintile or education); the only commonality is that indicators for multiple populations are being modeled within the same framework. 

The time period of interest can range any possible timespan, but usually includes a period beyond the most recent observation year in which case projections need to be made. Longer term projections may be of interest as well. For example, a common future year for projections is 2030, which is the target year for reaching the Sustainable Development Goals.

\begin{figure}[htbp]
    \centering
        \includegraphics[width=0.95\textwidth]{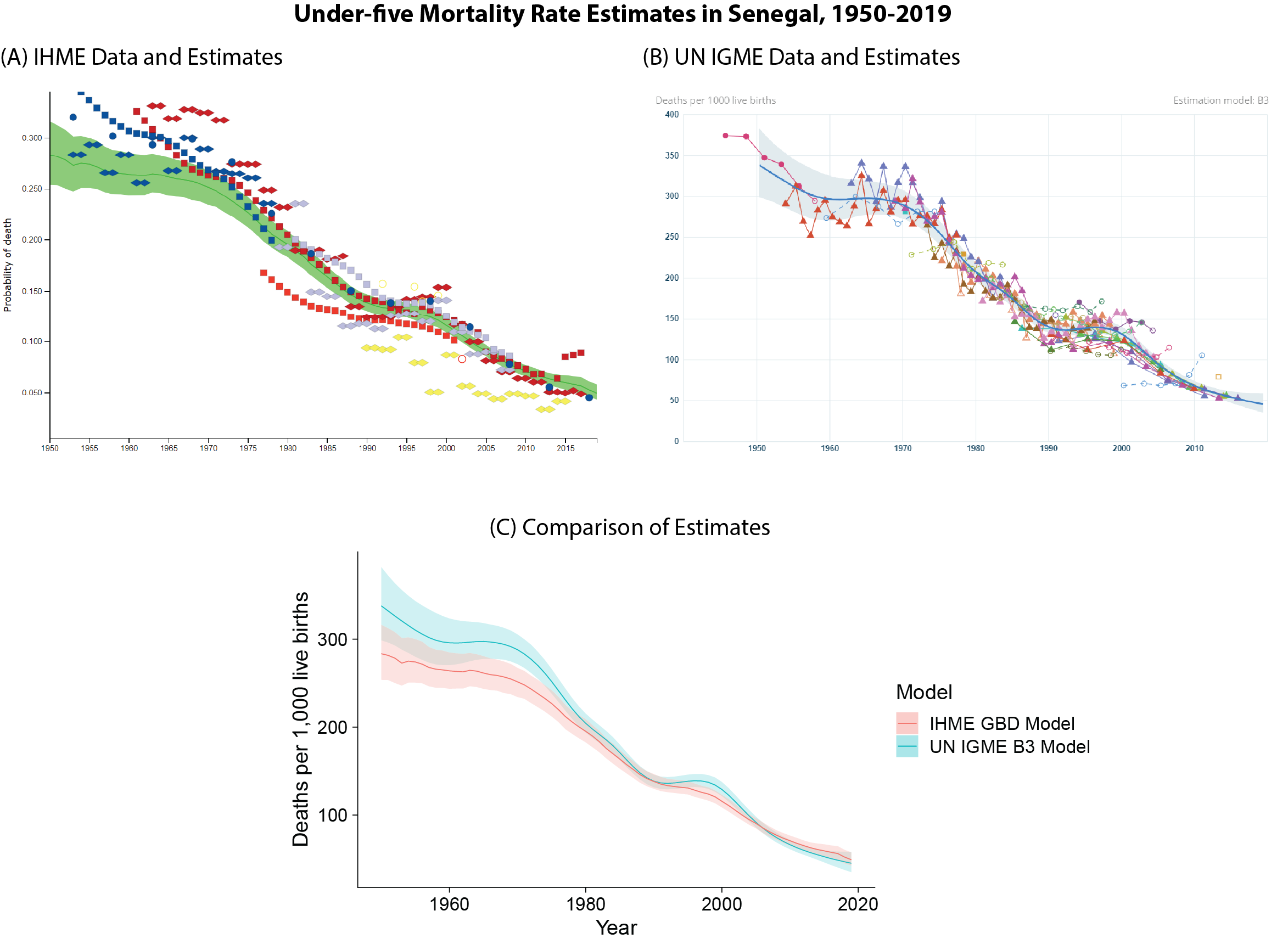}
        \caption{Comparison of estimates of the under-five mortality rate in Senegal from 1950-2019 from  the most recent estimates available from two models: the Global Burden of Disease (GBD) Model from the Institute of Health Metrics and Evaluation (IHME), and the B3 model from the UN Inter-agency Group for Child Mortality Estimation (IGME). While both models generate estimates of the same indicator, they have different data inclusion criteria and modeling assumptions, leading to diverging estimates. (A) Raw data and model estimates from the IHME Global Burden of Disease Study; source: \url{healthdata.org}, accessed 12/22/2020. (B) Raw data and model estimates from the UN IGME B3 model; source: \url{childmortality.org}, accessed 12/22/2020. (C) Comparison of estimates from each model on the same scale. }
    \label{fig:gbd_b3_comparison}
\end{figure}

In many contexts, there may not be a complete set of observations of the outcome of interest for all time points and populations. Additionally, data sources may be of differing quality, with varying levels of random error and systematic biases. The overall goal of a model is to use the available data to produce a full time series of the outcome for all populations of interest. Models need to be able to account for these errors and biases when producing estimates for years with observations, and be structured such that estimates can also be obtained for years without observations. Figure \ref{fig:gbd_b3_comparison} illustrates the problem by showing the data available for estimating the under-five mortality rate in Senegal from 1950-2019 and compares the resulting estimates from two models. 

Formally, let $t=1,\dots,T$ index the time period and $c=1,\dots,C$ index the populations of interest. The goal is to estimate the `true' value of the indicator in each population at every time point, which we denote $\eta_{c,t}$. The estimates should be probabilistic and accurately describe the uncertainty in the estimates of the true values. To inform the model there are observations $y_i$, $i=1,\dots,N$ of the indicator. Each of these observations is associated with a time point $t[i]$ and population $c[i]$. There may be multiple observations for the same population and time point, as well as populations with few or no observations. 

The goal in this paper is to propose a framework that unifies the modeling approaches used to describe the evolution of the underlying outcome of interest over time, $\eta_{c,t}$. A number of diverse modeling approaches have been applied to this problem. While all of the approaches we will consider generate probabilistic estimates, the way they go about it can be quite different. Without the additional structure provided by a model class, it can be difficult to compare across existing approaches. In the next section, we illustrate this difficulty by describing two existing models for the same indicator using their original notation, in which it is not obvious how to make direct comparisons between models.

\section{Case study: two models of under-5 mortality rate}
\label{section:case_study}

We first motivate the need for a model class to compare TMMPs by illustrating two important models that estimate the under-five mortality rate (U5MR) at the same geographic level and for the same time period. The U5MR, which is defined as the number of deaths to children under the age of five per 1,000 live births, is one of the most important indicators of child mortality, forming the basis of one of the Millennium Development Goals (MDGs) and now one of the Sustainable Development Goals (SDGs). In particular, SDG 3.2 stipulates that by 2030, all countries should reduce U5MR to at least as low as 25 per 1,000 live births. As a consequence of the focus on U5MR in the global health community, much work has been done on estimating and projecting trends in U5MR over time, as well as trying to understand these trends \citepalias{alkema_global_2014, dicker_global_2018, igme2020, li_changes_2019, wang2017global, you_global_2015}. 

In terms of producing reliable estimates and projections for U5MR, one of the main challenges occurs with trying to reconcile multiple, incomplete data sources in low-income countries, filling in gaps in the time series and giving reasonable bounds of uncertainty in estimates and projections. Two main groups have formulated models to estimate U5MR for all countries. The first is a group at the Institute of Health and Metrics Evaluation (IHME), as part of their work on the Global Burden of Disease (GBD) Study \citep{dicker_global_2018}. The second is the United Nations Interagency Group on Child Mortality Estimation (UN-IGME), whose model forms the basis of the official U5MR estimates produced by UNICEF \citep{alkema_global_2014}. The models produce differing estimates of U5MR: Figure \ref{fig:gbd_b3_comparison} compares the most recent model estimates covering 1950-2019 in Senegal as an example. Differences in the estimates could be due to differences in the data chosen for inclusion, differences in data processing, or differences in model assumptions \citep{alkema2012child}.

We will briefly describe these models following their original presentation to show how the use of different notational conventions makes them appear superficially different. Then, after we introduce our model class, we will return to these examples to show how casting them in our notation facilitates their comparison. In Section \ref{section:examples} we will expand the scope beyond these two models and show how several other existing models fit into the model class.

\subsection{Global Burden of Disease (GBD) Model}
The GBD model \citep{dicker_global_2018} has three modeling stages. The first two stages preprocess the observed data and the third stage provides the final estimates of U5MR. In the first stage, a nonlinear mixed effects regression model is fit to the U5MR observations, using lag-distributed income per capita (LDI), mean years of education for women of reproductive age (15-49 years), and HIV death rate in ages 0-4,  as covariates:
\begin{align*}
   y_i = \exp\left[ (\beta_1 + \gamma_{1,c[i]}) \log(x^{LDI}_{c[i], t[i]}) + (\beta_2 + \gamma_{2,c[i]})x^{educ}_{c[i], t[i]} + \gamma_{c[i]} + \gamma_{c[i], s[i]} + \alpha_{type[i]} \right] \\ 
   + (\beta_3 + \gamma_{3,c[i]}) x^{HIV}_{c[i], t[i]} + \epsilon_{c[i], t[i], s[i]}, \nonumber
\end{align*}
where $c[i]$, $t[i]$, and $type[i]$ are the country, year, and source type of observation $i$, respectively, and $\alpha_{type[i]}$ refers to a random effect to capture data-type-specific errors.\footnote{The equation supplied by the authors does not include a random intercept $\gamma_{3,c[i]}$ for the HIV covariate; we did not receive clarification from the corresponding author, so we took this omission to be an error and included the random effect.} Bias adjusted data points $y^{adjusted}_i$ were derived from the fitted coefficients and mixed effects according to the following formula:
\begin{align*}
    y^{adjusted}_i =& \exp\left[(\beta_1 + \gamma_{1,c[i]}) \log(x^{LDI}_{c[i], t[i]} + (\beta_2 + \gamma_{2,c[i]}) x^{educ}_{c[i], t[i]} + \gamma_{c[i]} + \gamma_{c[i], ref} + \alpha_{ref,c[i]} \right]  \nonumber  \\
    &+ (\beta_3 + \gamma_{3,c[i]) x^{HIV}_{c[i], t[i]}} + \epsilon_{c[i], t[i], s[i]},
\end{align*}
where $\gamma_{ref}$ and $\alpha_{ref}$ are the random effects from one or more reference data sources that are deemed to be more reliable than other sources by the study team. In the second modeling stage, predicted values are calculated by the formula
\begin{align*}
    y^{predicted}_i = \exp\left( \beta_1 \log(x^{LDI}_{c[i], t[i]}) + \beta_2 x^{educ}_{c[i], t[i]} + \alpha \right) + \beta_3 x^{HIV}_{c[i], t[i]},
\end{align*}
which is the same as the formula for the first stage, but without any random effects. The difference between $y^{predicted}$ and $y^{adjusted}$ was smoothed using a tricubic window to yield estimates in every country and year. Then a weighted combination of the smoothed differences and a local linear fit was used to yield estimated residuals (see \cite{dicker_global_2018} page 16-17 in Supplementary material 1; note that no mathematical description is provided).

In the third modeling stage, a Gaussian process is used to provide an additional level of temporal smoothing. The observed data are assumed to be normally distributed around the true value of the indicator on the log10-transformed scale:
\begin{align*}
    \log_{10} y_i \mid \eta_{c[i], t[i]} \sim N(\log_{10} \eta_{c[i], t[i]}, s_i^2).
\end{align*}
The observation variance $s_i^2$ is fixed depending on the data source. The true U5MR values are smoothed via a Gaussian Process, again on the log10-transformed scale:
\begin{align*}
     \log_{10} \bm{\eta}_{c} \mid \bm{\mu}_c, \bm{\Sigma}_c \sim MVN\left(\bm{\mu}_c, \bm{\Sigma}_c \right),
\end{align*}
where the mean of the process $\bm{\mu}_c$ are the log10-transformed $y^{predicted}$ plus the smoothed residuals from the second modeling stage, and $\bm{\Sigma}_c$ is a covariance matrix derived from a Matérn autocovariance function with fixed hyperparameters chosen based on a measure of data availability within each country.

\subsection{UN-IGME model}
The model used by the UN-IGME is referred to the `B3' model, which refers to the fact that the model is a Bayesian, bias-adjusted B-Splines model (\cite{alkema_global_2014}). This model set-up also assumes that the observed data are normally distributed around a true value:
\begin{align*}
    \log y_i &= \log(\eta_{c[i], t[i]}) + \delta_i,
\end{align*}
where error $\delta_i$ is normally distributed with mean and variance depending on the data source. 

For complete vital registration systems, the error term is normally distributed with zero mean and a fixed stochastic variance $v_i^2$:
\begin{align*} 
    \delta_i \sim N(0, v_i^2).
\end{align*}
For survey observations or data from other sources, the sampling variance is decomposed into a fixed sampling error and estimated non-sampling error. Systematic biases are also estimated, either as a linear trend for data sources with multiple observations or as an offset otherwise:
\begin{align*}
    \delta_i \mid \Psi_i, \Sigma_i \sim N(\Psi_i, \Omega_i + v_i^2),
\end{align*}
where $\Psi_i$ is a systematic bias parameter and $\Omega_i$ is an estimated non-sampling error. 

The true U5MR values $\eta_{c,t}$ are smoothed via a cubic B-spline model:
\begin{align*}
    \log(\eta_{c,t}) = \sum_{k=1}^{K_c} b_{c,k}(t)\alpha_{c,k},
\end{align*}
where $K_c$ are the number of knots in country $c$, $b_{c,k}$ is the $k$th spline value in country $c$, and $\alpha_{c,k}$ is the $k$th spline coefficient. The spline coefficients during the observation period are parameterized as following a linear trend with random walk deviations away from that trend:
\begin{align*}
\alpha_{c,k} &= \lambda_{c,0} + \lambda_{c,1}(k - K_c/2)+[\bm{D}_{K_c}'(\bm{D}_{K_c}\bm{D}_{K_c}')^{-1}\bm{\varepsilon}_c]_k
\end{align*}
where  $\lambda_{c,0}$ and $\lambda_{c,1}$ are the unknown level and slope parameters for the spline coefficients in country $c$ and parameter vector $\bm{\varepsilon}_c = (\varepsilon_{c,1},\hdots, \varepsilon_{c,Q_c})'$ contains the $Q_c = K_c-2$ second-order differences in the spline coefficients, $\varepsilon_{c,q} = (\alpha_{c,q+2} - \alpha_{c,q+1}) - (\alpha_{c,q+1} - \alpha_{c,q})$ for $q=1, \hdots, Q_c$; $[\bm{D}_{K_c}'(\bm{D}_{K_c}\bm{D}_{K_c}')^{-1}\bm{\varepsilon}_c]_k$ refers to the $k$-th element of vector $\bm{D}_{K_c}'(\bm{D}_{K_c}\bm{D}_{K_c}')^{-1}\bm{\varepsilon}_c$, with known difference matrix $\bm{D}_{K_c}$ (defined by $D_{K_c,i,i} = D_{K_c,i,i+2} = 1$, $D_{K_c,i,i+1} =-2$ and  $D_{K_c,i,j} = 0$ otherwise).  Vague priors were used for the $\lambda_{c,0}$'s and the $\lambda_{c,1}$'s. Second-order differences are penalized by imposing
\begin{align*}
\varepsilon_{c,q}|\sigma_c^2 &\sim N(0, \sigma_c^2),\text{ for } q = 1, \hdots, Q_c,
\end{align*}
where variance $\sigma_c^2$ determines the extent of smoothing; a smaller variance corresponds to smoother trajectories. A multilevel model is placed on the smoothing parameters, i.e. the standard deviation of $\varepsilon_{c,q}$, to share information between countries:
\begin{align*}
\log(\sigma_c)|\chi,\varphi_{\sigma}^2  &\sim N(\chi, \varphi_{\sigma}^2),
\end{align*}
where $\chi$ and $\varphi_{\sigma}^2$ refer to the mean and variance of the log-transformed standard deviations. 

UN IGME uses B3 to produce estimates of U5MR up to the most recent year. A logarithmic pooling approach was used to produce estimates past the most recent observation year to combine country-specific posterior predictive distributions (PPDs) for changes in spline coefficients with a global PPD. This procedure was applied to modify the PPDs for $\alpha_{c,k}$ for $k=K_c, K_c+1, \hdots, P_c$, where $K_c$ and $P_c$ refer to the indices of the most recent splines in the observation and projection periods respectively. The approach is summarized (leaving out indices referring to posterior samples) as follows: 
\begin{align*}
\varepsilon_{c,K_c+a}|\Gamma_{c,K_c+a},\Theta_{c,K_c+a} &\sim  N\left(\Gamma_{c,K_c+a}, \Theta_{c,K_c+a}\right), \text{ for } a \geq 0,
\end{align*}
where
\begin{align*}
\Gamma_{c,k} &= W\cdot G+ (1-W)\cdot \varepsilon_{c,k-1},\\
\Theta_{c,k} &= W\cdot V + (1-W)\cdot \Theta_{c,k-1},
\end{align*}
with $G$ and $V$ equal to the median and variance of the estimates of past 2nd-order differences $\hat{\varepsilon}_{1:C,1:K_c}$'s respectively, and $\Theta_{c,K_c-1} = \sigma_c^2$. The overall pooling weight $0\leq W \leq 1$ was chosen through an out-of-sample validation exercise.

\subsection{Comparing models}
Based on these overviews we see immediately that each of the two models take diverging approaches to modeling U5MR. They make different choices about use of covariates and how to impose smoothness on the final estimates. For example, while the GBD model uses covariates derived from other data sources, such as LDI and education levels, the B3 model relies solely on data-driven trends that are smoothed using penalized B-splines. However, it is less clear how to systematically compare the two. The three modeling stages of the GBD model make it difficult to compare directly to the integrated modeling approach of the B3 model. We would like to be able to compare the assumptions each models makes about how the true value of the indicator evolves, its relationship to the observed data, and how information is shared between countries. The goal of our proposed model class for TMMPs is to aid in making such direct comparisons.

Throughout the paper, in addition to illustrating how the two U5MR models can be re-expressed within the TMMP framework, we will also call on other example models from the demographic and health estimation literature, including models to estimate maternal mortality \citep{alkema_bayesian_2017}, neonatal mortality \citep{alexander_global_2018}, contraceptive use \citep{cahill_modern_2018} and age-specific mortality rates \citep{alexander_flexible_2017}. Each of these examples highlights important aspects of the general TMMP formalization in different settings. 

\section{Model class: TMMPs}
\label{section:framework_overview}

In this section we present a general notation for describing a modeling framework for TMMPs. As a general rule, we use Greek letters for unknown parameters and Latin letters for fixed values. For example, $\sigma^2$ for an unknown variance to be estimated, while $s^2$ for a fixed sampling error variance. Bold face is used to denote matrices.

\begin{figure}[htbp]
    \centering
        \includegraphics[width=\textwidth, height=0.55\textwidth]{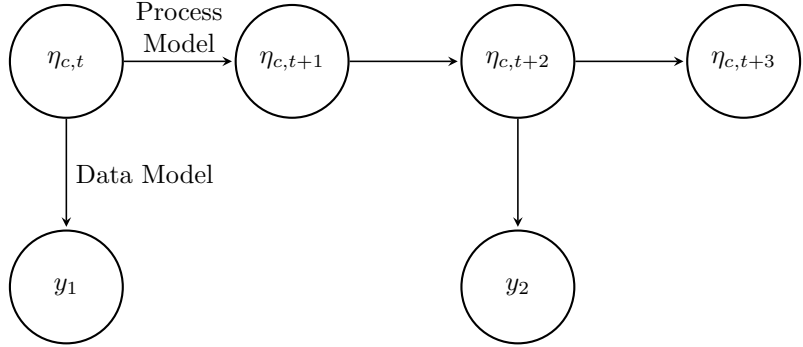}
        \caption{The model class distinguishes between the true values of an indicator $\eta_{c,t}$ and the noisy observed data $y_i$. The process model describes the evolution of the true values, and the data model describes how the observed data are generated from the true values. This structure handles missing data naturally: the latent trend is modeled for all time points by the process model, and it is possible that only some time points have observed values. In this example, observed data only exist for times $t$ and $t+2$. \label{fig:model_diagram}} 
\end{figure}

The framework is organized around a distinction between observed data and latent trends. The observed data are noisy, possibly biased observations of the latent trend. Let $\eta_{c,t}$ be the latent true value of the indicator in country $c$ at time $t$, which we emphasise is an unobserved parameter that needs to be estimated. Under this notation, modeling requires specifying a relationship between the observed $y_i$ and latent $\eta_{c,t}$, which we refer to as the data model (also often referred to in the literature as the likelihood), and describing the temporal evolution of the latent trend, which we refer to as the process model (Figure \ref{fig:model_diagram}).

Our proposed model class draws on the rich literature on models involving latent structures and measurement error. Measurement error models acknowledge uncertainty and bias in observed data \cite{carroll2006measurement}, as we do in our model class. State space models also make a distinction between noisy observations and an underlying latent process \citep{chatfield2019analysis}. There is a vast literature on flexible and data-driven modeling techniques, especially for time-series, which we draw on in our framework \citep{west2006bayesian}. Our model class tailors these broad existing approaches to the context of demographic and health indicators by providing an overarching structure. This provides a principled way of combining the ideas behind measurement error models, state space models, and smoothing models into interpretable models suited for the application of interest.


The process model describes the evolution of the $\eta$s. The process model is separated into covariate, systematic, offset, and smoothing components:
\begin{align} \label{eq-processmodel}
    g_1(\eta_{c,t}) = g_2({X}_{c,t}, \bm{\beta}_c) + g_3(t, \eta_{c,s \neq t}, \bm{\alpha}_c) + a_{c,t} + {\epsilon}_{c,t},
\end{align}
where
\begin{itemize}

\item $g_1$ is a transformation of $\eta_{c,t}$.
\item $g_2({X}_{c,t}, \bm{\beta}_c)$ is the covariate component, a function of covariates $X_{c,t}$ and parameter vector $\bm{\beta}_c$.
\item $g_3(t,\eta_{c, s \neq t}, \bm{\alpha}_c)$ refers to a systematic temporal component, a mathematical function which describes a parametric trend over time, that may depend on prior or future values ${\eta}_{c, s \neq t}$.
\item $a_{c,t}$ is a fixed offset which may be used to incorporate information from models estimated in a separate process.
\item ${\epsilon}_{c,t}$ is a deviation term to allow for data-driven deviations from the trends described by $g_2$, $g_3$, and the fixed offset. The vector $\bm{\epsilon}_c$ is decomposed as $\bm{\epsilon_c} = \bm{B}\bm{\delta}_c$, where $\bm{B}$ is a full rank matrix that can be used to transform $\bm{\epsilon}$ to a lower-dimensional space, e.g. with a B-spline basis approach. The term satisfies $E(\triangle_r {\epsilon}_{c,t})= 0$ for some $r\geq 0$, and, if $r > 0$, 
\begin{align*}
\sum_{k \in \mathcal{K}_{d,c}} \triangle_d \delta_{c,k} = 0, \text {for } 0 \leq d \leq (r-1),
\end{align*}
for set of indices $\mathcal{K}_{d,c}$. 
\end{itemize}

The process model controls the behavior of the model particularly in projections and data sparse time periods. 
The chosen form of the process model affects not only the central estimate of projections and periods with missing data, but also the estimated uncertainty around those point estimates. While the process model in Eq.~\ref{eq-processmodel} is specified as a marginal model for one indicator $\eta$ at time $t$, correlation structures over time, space, or other dimensions can be incorporated through the specification of the process model components. We will consider each component of the process model separately with attention to their contribution to the overall behavior of a model. 

The data model connects the latent trends of the process model to the observed data. The data model should allow for different response distributions, data transformation, and biases depending on the observation. This flexibility would allow us to model differing sampling errors, non-sampling errors, and systematic biases depending on the data source of each observation. For example, a logit-normal data can be expressed as:
\begin{align*}
    \mathrm{logit}(y_i) \mid \eta_{c[i], t[i]}, \sigma_i^2 \sim N(\eta_{c[i], t[i]}, \sigma_i^2),
\end{align*}
where $\sigma_i^2$ is the estimated variance of the observations. Extensions to this simple data model could include explicitly incorporating biases and decomposing the variance into sampling and non-sampling error. We leave developing a notation for data models to future work, focusing instead in this article on the details of the process model.

\subsection{Covariates}

In some cases, particularly in data sparse settings, it might make sense to include covariates in the process model, such that the indicator of interest can be modeled as a function of related factors which may have more data available. For example, there is often a strong association observed between health indicators and measures of the economy or wealth, such as a country's gross domestic product (GDP). While information on specific health indicators may be sparse, estimates of GDP are widely available. The addition of covariates and the modeling of such associations can be thought of in the same way as a standard generalized linear model framework, with other aspects of the process model adding greater complexity and allowing for non-linear trends to be captured. 

Covariates can be included in the process model via the regression function $g_2(\bm{X}_{c,t}, \bm{\beta}_c)$, where $\bm{\beta}_c$ are country-level parameters. The choice of covariates to include may be informed by exploratory data analyses or substantive knowledge of the covariates that are associated or causally related to the outcome of interest. The covariate-based component is useful for providing expected trends of indicators in projections and data-sparse settings.  

Table~\ref{tab-cov} gives examples of regression functions used in existing models of global health indicators. For example, a model to estimate maternal mortality ratios (MMRs) discussed in \cite{alkema_bayesian_2017} uses Gross Domestic Product (GDP), General Fertility Rate (GFR), and percentage of births with a skilled attendant present (SAB) as covariates for estimating the log-transformed proportion of non-AIDS maternal deaths. The neonatal model discussed in \cite{alexander_global_2018} uses U5MR as a covariate, where U5MR is used as a predictor for the ratio of NMR to (NMR - U5MR), that is, neonatal to other child mortality. The inclusion of U5MR as a covariate is based on the strong demographic relationship that in general, as child mortality decreases, the share of deaths that are neonatal increases. The form of this relationship is piecewise linear, based on the results of an exploratory data analysis. The GBD U5MR model of \cite{dicker_global_2018} uses a non-linear regression function incorporating lag-distributed income per capita (LDI), mean years of education for women age 15-49 (EDU), and the HIV death rate in ages 0-4 (HIV). Finally, the age-specific mortality model of \cite{alexander_flexible_2017} uses the principle components of mortality schedules as covariates.

\begin{table}[htbp]
\begin{tabular}{
    | P{0.2\textwidth}
    | p{0.2\textwidth}
    | p{0.05\textwidth}
    | P{0.28\textwidth}
    | p{0.2\textwidth}
    |
} \hline
Indicator    & $\eta_{c,t}$ & $g_1(\cdot)$ & $g_2(\cdot)$ & covariates \\ 

\hline

Maternal mortality ratio \citep{alkema_bayesian_2017} & proportion of non-AIDS deaths that are maternal & $\log$ & $\beta_{c,0} + \sum_k X_{c,t,k}\beta_k$ & log(GDP), log(GFR), SAB    \\ 

\hline 
Neonatal mortality rate \citep{alexander_global_2018}   & NMR/(U5MR-NMR) & logit & $\beta_{0,c} + \beta_1 \cdot \left( \log(X_{c,t}) - \log(\beta_2) \right)\bm{1}_{[X_{c,t} > \beta_2]}$ & U5MR  \\

\hline

Sub-national mortality \citep{alexander_flexible_2017} & age-specific mortality & log & $\sum_k X_{c,t,k} \beta_{c, t, k} $ & principal components of mortality schedule \\

\hline

  U5MR \citep{dicker_global_2018} & crisis-free U5MR  & $\log_{10}$ & $\exp[ \beta_{c,1} \cdot \log(X^{LDI}_{c, t}) + \beta_{2,c} \cdot X^{EDU}_{c, t} + \beta_{3,c}] + \beta_{4,c} x^{HIV}_{c, t}$  & LDI, EDU, HIV
   
  \\ \hline 
  
\end{tabular}
\caption{Selected examples of covariate functions.}\label{tab-cov}
\end{table}

\subsection{Systematic component}
We may expect the trend over time of some demographic and health indicators to follow a certain path that can be expressed as a parametric function. For example, demographic transition theory suggests that as a country's fertility rate declines, it will first decline rapidly, and then decelerate, and eventually plateau at a certain level \citep{kirk1996demographic}. A similar assumption about the shape of the rate of change can be made for contraceptive prevalence over time. Such assumptions can be incorporated by a mathematical function $g_3(t, \eta_{c,s \neq t}, \bm{\alpha}_c)$ which encodes a parametric trend depending on a set of parameters $\bm{\alpha}_c$. The form of this function may be informed by prior knowledge of the mechanisms that drive the evolution of the indicator of interest. 

Table~\ref{tab-str} gives examples of systematic components used in a selection of existing models. For example, the process model of \cite{cahill_modern_2018} describes the total contraceptive use rate in a country over time. Based on a modeling assumption that a country's total contraceptive use rate will increase slowly, speed up, then taper off as it moves through the contraceptive use transition, their model includes a systematic component which captures an expected rate of change based on a logistic growth curve. Note that this type of systematic component, which defines the level of an indicator by calculating its change from its previous value, results in increasing variance in projections. This is discussed further in Section \ref{section:fpem-example}.

\begin{table}[htbp]
\begin{tabular} {
    | P{0.18\linewidth}
    | p{0.8\linewidth} |
}
    
    \hline
    
    Indicator & $g_3(\cdot)$ \\
    
    \hline
    
    Total contraceptive use \citep{cahill_modern_2018} & 
        Logistic curve: \newline 
        $g_3(\cdot) = \Omega_c$ when $t = t^*$, and for $t>t^*$: \newline
        $g_3(\cdot) =\mathrm{logit}\left(\eta_{c,t-1}\right) + \iota_{c,t}$
        \newline
        $ = \begin{cases}
        \mathrm{logit}\left( \tilde{P}_c \cdot \mathrm{logit}^{-1}\left( \mathrm{logit}\left(\frac{\eta_{c, t-1}}{\tilde{P}_c}  \right) + \omega_c \right) \right), & \text{ when } \eta_{c,t-1} < \tilde{P}_c \\
       \mathrm{logit}\left(\eta_{c,t-1}\right), & \text{otherwise,}
    \end{cases}$
    \newline
        where $\bm{\alpha}_c = \left\{ \tilde{P}_c, \omega_c, \Omega_c \right\}$. \\

    \hline 
  
    Inflation in the sex ratio at birth \citep{chao_systematic_2019} &
      Trapezoid function: \newline
      $\begin{cases}
          \xi_c(t - \gamma_{0, c}) \slash \lambda_{1, c}, & \gamma_{0, c} < t < \gamma_{c,1} \\
          \xi_c, & \gamma_{1, c} < t < \gamma_{2, c} \\
          \xi_c - \xi_c(t - \gamma_{2, c}) \slash \lambda_{3, c}, & t < \gamma_{2,c} < t < \gamma_{3,c} \\
          0, & \text{otherwise}
      \end{cases}$
      
      where  $\bm{\alpha}_c = \left\{ \xi_c, \gamma_{0, c}, \lambda_{c,1}, \lambda_{2,c} \right\}$, $\gamma_{c,1} = \gamma_{c,0} + \lambda_{c,1}$, $\gamma_{2,c} = \gamma_{c,1} + \lambda_{2,c}$, and $\gamma_{3,c} = \gamma_{2,c} + \lambda_{3,c}$. \\
      
      \hline
\end{tabular}
\caption{Selected examples of systematic components.}\label{tab-str}
\end{table}

\subsection{Offset}

Some models incorporate information from separate modeling stages into their process model. The TMMP model class accounts for this through the $a_{c,t}$ term in the process model. Alternatively, the offset could be integrated into the covariate component, akin to the use of offsets in generalized linear models. Here we chose to separate it into a separate component to make it clearer when models incorporate fixed external information besides covariates. For example, the GBD U5MR model uses an offset which is derived from the smoothed residuals from a regression model estimated separately and serves to adjust the output of the covariate component.

\subsection{Smoothing component}

The final part of the process model, the smoothing component, captures trends that are not explained by the systematic or covariate components. This component allows for trends over time to be driven by patterns observed in the data. To prevent overfitting, a model is placed on the deviations to enforce some degree of smoothness in the model fit. 
We define a smoothing model as the distribution placed on the joint vector of all deviations from a country, $\bm{\epsilon}_c = [\epsilon_1, \cdots, \epsilon_T]$. The smoothing model is defined as follows: 
\begin{align}
    \bm{\epsilon}_c &= \bm{B}_c \bm{\delta}_c, \label{eq-smoother1}
\end{align}
where $\bm{B}_c$ is a $T \times K_c$, $K_c \leq T$ full rank matrix and $\bm{\delta}_c = [\delta_{c,1}, \delta_{c,2}, \hdots, \delta_{c,K_c}]$ is a vector of parameters. 

We define the parameter vector $\bm{\delta}_c$ or its differenced version to be normally distributed: for $r \geq 0$
\begin{align}
    \triangle_r  \bm{\delta}_c \mid \bm{\Sigma}_c &\sim N(\bm{0}, \bm{\Sigma}_c),\label{eq-smoother2}
\end{align}
where $\triangle_r$ is the difference operator with $\triangle_1 \bm{\delta} = [\delta_2 - \delta_1, \hdots, \delta_k - \delta_{K_c-1}]$ and $\triangle_r\bm{\delta} = \triangle_1^r\bm{\delta}$. When $r>0$, to ensure the model is generative we need to introduce extra model structure to anchor the overall level of the deviations. 
Formally, we require 
\begin{align}
\sum_{k \in \mathcal{K}_{d,c}} \triangle_d \delta_{c,k} &= 0, \text {for } 0 \leq d \leq (r-1),\label{eq-smoother4}
\end{align}
for set of indexes $\mathcal{K}_{d,c}$.

The covariance matrix $\bm{\Sigma}_c$ is specified via an autocovariance function $s$, which may depend on hyperparameters. We restrict the set of autocovariance functions to those that can be expressed as a function only of the absolute distance between $t_1$ and $t_2$:
\begin{align}
\Sigma_{c, t_1, t_2} &=  s(t_1, t_2) = s^*(|t_1 - t_2|). \label{eq-smoother3}
\end{align}
We further require that the covariance kernel goes to zero as $|t_1 - t_2|$ goes to infinity. Commonly used autocovariance functions in this situation are the autocovariance associated with an autoregressive process of order 1 (AR(1) process), the squared exponential covariance function, and the Matérn covariance function, which have the following forms:
\begin{align}
    s_{\mathrm{AR1}}(t_1, t_2) &= \kappa^2 \rho^{|t_1 - t_2|}, \\
    s_{\mathrm{SE}}(t_1, t_2) &= \kappa^2 \exp\left(-\frac{(t_1 - t_2)^2}{2\ell^2}\right), \\
    s_{\mathrm{Mat\acute{e}rn}}(t_1, t_2) &= \kappa^2 \frac{2^{1-\nu}}{\Gamma(\nu)} \left( \sqrt{2\nu} \frac{|t_1 - t_2|}{\ell} \right)^\nu K_\nu \left(\sqrt{2\nu} \frac{|t_1 - t_2|}{\ell} \right). \label{eq-matern}
\end{align}

Smoothing models that include a transformation (that is, $\bm{B}_c \neq \bm{I}$) include B-spline approaches \citep{eilers1996flexible}. The transformation $\bm{B}_c$ can be used to lower the dimension of the smoothing model; for B-splines, for example, knots could be placed on a sparse grid in order to reduce the number of smoothing parameters to estimate. In this setup the smoothing parameters $\bm{\delta}_c$ are the coefficients of the spline basis functions. Since the smoothing model is the linear combination of differentiable functions, the smoothing model itself will be differentiable, which may not be the case for other choices of the smoothing model. Differentiability may be a desirable property because it implies a degree of smoothness on the smoothing model.

This definition of smoothing models as summarized in Equations~\ref{eq-smoother1}, \ref{eq-smoother2}, \ref{eq-smoother3}, and \ref{eq-smoother4} encapsulates main methods of smoothing, including ARIMA processes, random walks, Gaussian processes, and B-splines, as illustrated in Table~\ref{tab-smooth}. 

As mentioned earlier, for non-stationary models with $r>0$, we require the constraint
\begin{align*}
\sum_{t \in \mathcal{K}_{d,c}} \triangle_d \delta_{c,k} &= 0, \text {for } 0 \leq d \leq (r-1),
\end{align*}
for set of indexes $\mathcal{K}_{d,c}$. This requirement makes the model identifiable by anchoring the sum of the differenced process at zero over the set $\mathcal{K}_{d,c}$. If the smoothing model does not perform any dimensionality reduction by its choice of $\bm{B}$ then $\mathcal{K}_{d,c}$ is interpretable as the set of years for which the differenced process sums to zero. Common choices include $\mathcal{K}_{d,c}$ fixed at a specific reference year, i.e., as used for family planning estimation \citep{cahill_modern_2018} and maternal mortality \citep{alkema_bayesian_2017}.  $\mathcal{K}_{d,c}$ might also be chosen to span the observation period of the country of interest. For the models used in neonatal and child  mortality estimation, the sum of the spline coefficients are constrained to sum to zero, corresponding to $\mathcal{K}_{d,c} = \left\{1, \dots, K_c\right\}$ where $K_c$ are the number of splines in country $c$ \citep{alexander_global_2018, alkema_global_2014}. 

The behavior of the smoothing model may influence the trends and uncertainty of projections. Appendix~B provides further detail on stationary and non-stationary models, including how they behave in projections.

\begin{table}[hbtp]
\centering
\begin{tabular}{
    | P{0.25\textwidth}
    | p{0.18\textwidth}
    | p{0.2\textwidth}
    | c
    | p{0.2\textwidth} 
    |
} \hline
    Indicator    & $\bm{B}$  &  $k(t_1, t_2)$ & $r$ & $\mathcal{K}_{d,c}$ \\ 
    
    \hline
    
    Maternal mortality ratio \citep{alkema_global_2016} & $\bm{B} = \bm{I}$ & ARMA(1, 1)
    & 1 & $\left\{ 1990 \right\}$  \\ 
    
    \hline
    
    U5MR \citep{alkema_global_2014} & $B_{c,t,k} = b_{c,k}(t)$= \newline cubic B-splines, knots 2.5 years & indep. $s(t_1, t_2) = \sigma^2 1(t_1 = t_2)$& 2  & $\mathcal{K}_{0,c} = \left\{ k^* \right\}$, \newline $\mathcal{K}_{1,c} = \left\{ 2, \cdots, K_c \right\}$
    \\ 
    
    \hline
    
  U5MR \citep{dicker_global_2018} & $\bm{B} = \bm{I}$ & Matérn, see Eq.\ref{eq-matern} & 0 & $\cdot$ \\ 
    
    \hline
\end{tabular}
\caption{Selected examples of smoothing processes.}\label{tab-smooth}
\end{table}

\subsection{Projections}

It is commonly of interest to produce estimates beyond the last observed data points. Suppose the last data point in a country is observed at $T$, and we would like to project the value of the indicator to time $T^*$. Using the estimation process model for projections is taken as default in the TMMP framework. In particular, based on the estimation model, $\eta_{c,t}$ can be estimated for $1 < t <= T^*$ as part of the joint estimation process of the entire model. Alternatively, $\eta_{c,t}$ can be estimated up until $T$, the last available data point, and the estimation process model can be used afterwards to generate projections from $T$ to $T^*$. We make explicit how projections come about for TMMP models at the end of this section. 

In some modeling approaches, the projection model used for $\eta_{c,t>T}$ may differ from the estimation model. For example, in the B3 model, a pooling approach is used to  control the variance in the projections. For such models, the projection model used needs to specified separately, either as a new TMMP model, or in terms of how the model differs from the estimation model. 

\paragraph{Default projections obtained from TMMP models}
For TMMP models, the smoothing component up to year $T^*$ is given by
\begin{align}
    \bm{\epsilon}_c = \bm{B}_c \bm{\delta}_c,
\end{align}
where $\bm{\epsilon}_c$ is a $T \times 1$ vector, $\bm{B}_c$ is a $T \times K_c$ matrix, and $\bm{\delta}_c$ is a $K_c \times 1$ vector. To extend the smoothing component to the end of the projection period $T^*$, we define an extended vector of smoothing terms $\bm{\epsilon}^*_c$ with
\begin{align}
    \bm{\epsilon}^*_c = \bm{B}^*_c \bm{\delta}^*_c, \label{eq-smootherextended}
\end{align}
where $\bm{\epsilon}^*_c$ is a $T^* \times 1$ vector, $\bm{B}_c$ is a $T^* \times K_c^*$ matrix, and $\bm{\delta}_c$ is a $K_c^* \times 1$ vector. 
Note that in a B-spline setup, this likely includes adding extra knots to cover the projection period, which corresponds to adding additional columns to $\bm{B}^*$. When the projection uses the estimation process model, we have that
\begin{align}
    \triangle_r \bm{\delta}^*_c \mid \bm{\Sigma}^*_c \sim N\left(\bm{0}, \bm{\Sigma}^*_c \right). \label{eq-deltaconditional}
\end{align}
The distribution of the (differenced) smoothing terms in the projection period follow from the conditional distribution $\triangle_r \delta^*_{c, t > T} \mid \bm{\delta}_c, \bm{\Sigma}_c$, which has a closed form. 

For TMPPs with smoothers, the TMMP projection process model defaults to $r$th-order differenced estimation process model. For example, if the smoothing model has $r=0$, then the projection model is given directly by the estimation model with the extended smoothing term as per Eq.~\ref{eq-smootherextended}. If $r=1$, then the projections are based on the first order differences of the process model. To illustrate, consider a process model that includes a smoothing component with $r=1$. The process model is written
\begin{align}
    \eta_{c,t} = g_2(X_{c,t}, \bm{\beta}_c) + g_3(t, \eta_{c,s \neq t, \bm{\alpha}_c}) + \epsilon_{c,t},
\end{align}
where $\bm{\epsilon}_c = \bm{B}_c \bm{\delta}_c$. 
The first order difference of the process model is given by
\begin{align}
    \eta_{c,t} - \eta_{c,t-1} =& (g_2(X_{c,t}, \bm{\beta}_c) - g_2(X_{c,t-1}, \bm{\beta}_c))  \nonumber \\
    & + (g_3(t, \eta_{c,s\neq t}, \bm{\alpha}_c) - g_3(t - 1, \eta_{c, s \neq t- 1}, \bm{\alpha}_c)   \nonumber \\
    & + (\epsilon_{c,t} - \epsilon_{c,t-1}).
\end{align}
The conditional distribution of the first order differences $(\epsilon_{c,t} - \epsilon_{c,t-1})$ in the projection period then follows from Equations~\ref{eq-smootherextended} and \ref{eq-deltaconditional}.

\subsection{Parameter estimation}
\label{section:hierarchical_modeling}
The TMMP framework requires estimating many country-level parameters. In the process model the systematic trend depends on $\bm{\alpha}_c$, the covariate function depends on $\bm{\beta}_c$, and the smoothing model may have country-specific hyperparameters. To complete the specification of a model, we need to define how these parameters will be estimated. In this section we describe several approaches, including fixing parameters, use of informative priors, and hierarchical modeling, and present examples of how they have been used in existing models. These approaches can be easily combined within a model: for example, informative priors might be set for some parameters while hierarchical distributions are used for others. 
Documentation of the assumptions made is important to make explicit the sharing of information, for example across populations, and the use of external information to inform the estimates.  

\subsubsection{Fixed parameters}
Parameters can be fixed by the modeler to set values based on substantive knowledge, convenience, or results from separate models. For example, \cite{dicker_global_2018} set the country level hyperparameters of their smoothing model to fixed values depending on a measure of data availability in each country.
They also use fixed parameters in the covariate term in the process model. The regression coefficients are obtained from fitting a mixed effects regression model to a global data set. The offset terms $a_{c,t}$ are obtained from smoothing the residuals of that regression.  

\subsubsection{Priors}
Prior distributions can be applied to parameters to inform estimation. When substantive information is available for parameters then informative priors can be used. Vague priors can be used when there is little or no prior knowledge to inform the parameter values. The models considered in this paper as examples use vague priors for most model hyperparameters. We note that informative priors have been used more often for data bias and measurement error parameters in data models, a discussion of which we leave to future work.

\subsubsection{Hierarchical modeling}
Hierarchical modeling is a general approach that can be used for parameter estimation when it is desirable to share information between units. These parameters can be estimated in different ways according to how the modeler wants to share information between countries. At one extreme, the parameters could be estimated independently for each country. This encodes an assumption that the parameter in one country is unrelated to the parameter in another country. Another extreme is to have every country share the same set of parameters, which is appropriate for modeling global level patterns that are shared across all countries. In between these two options is an intermediary choice in which information is shared between countries via hierarchical modeling. By using hierarchical modeling, the parameter estimates in one country can be informed by trends in similar countries. Sharing information can be especially useful in cases where countries have limited or no data available, but the modeler still wishes to generate estimates. 

To ease comparison between models we introduce a basic notation for hierarchical models which encodes distributional assumptions and the levels and groupings of countries. For a country-level parameter $\gamma_c$, which may be defined on a transformed scale and can belong to any of the process model components, a hierarchical model can be written as
\begin{align*}
    \gamma_c \mid \gamma_{r[c]}^{(region)}, \sigma^{(region)2}_{\gamma} \sim \pi(\gamma_{r[c]}^{(region)}, \sigma^{(region)2}_{\gamma}),
\end{align*}
where $\pi$ represents a probability distribution. The parameters $\gamma_{r[c]}^{(region)}$ are the group-level means, where $r[c]$ indexes the group that country $c$ belongs to. The parameter $\sigma^{(region)2}_{\gamma}$ is the variance of the country level parameters around their group means. This setup can be extended recursively to allow for deeper hierarchies, where the group $r[c]$ can be a member of a higher-level group itself, and so forth. Specifying a hierarchical model therefore requires defining the distribution $\pi$, the number of levels in the hierarchy, and the hierarchical groupings. A common setup is to take $\pi$ to be the normal distribution and to have only one level of hierarchy, in which case $r[c]=1$ for all $c$ and $\gamma^{(region)}_1$ represents a global mean:

\begin{align*}
    \gamma_c \mid \gamma^{(world)}, \sigma_\gamma^2 \sim N(\gamma^{(world)}, \sigma^2_{\gamma}).
\end{align*}
The use of hierarchical modeling in several models is compared in Table \ref{tab:hierarchical_modeling}, giving their choices for distribution $\pi$ and the number of levels and hierarchical groupings. 

\begin{table}[hbtp]
\begin{tabular}{
    | P{0.25\textwidth}
    | P{0.25\textwidth}
    | p{0.1\textwidth}
    | p{0.05\textwidth}
    | p{0.3\textwidth} |
} \hline
    Indicator    & Hierarchical Models & $\pi$ & Levels & Groupings \\
    
    \hline
    
    \multirow{ 4}{0.25\textwidth}{Total contraceptive use \\ \citep{cahill_modern_2018}} & asymptote $\tilde{P}$ & normal & 1 & countries within world \\
    
    & rate $\omega_c$ & normal & 3 & countries within sub-region, \newline region, world \\
    
    & timing $\Omega_c$, developing countries & normal & 3 & countries within sub-region, \newline region, world \\
    
    & timing $\Omega_c$, developed countries & normal & 1 & countries within world \\
    
    \hline
    
    Maternal mortality \citep{alkema_bayesian_2017} & smoothing parameters $\lambda_c$ & truncated normal & 1 & countries within world \\
    
    \hline
    
    Age-specific mortality \citep{alexander_flexible_2017} & regression coefficients $\beta_{k, a}$ & normal & 1 & counties within state \\
    
    \hline
\end{tabular}
\caption{Hierarchical modeling in selected models. }\label{tab:hierarchical_modeling}
\end{table}

\subsection{Recommendations for TMMP reporting guidelines}
We recommend that the writing of model assumptions for TMMPs, including those related to parameter estimation, in a standardized way is considered for a future version of GATHER, referring to Guidelines for Accurate and Transparent Health Estimates Reporting \citep{stevens2016guidelines}. We provided a template for doing so in Appendix A and illustrate its use for the case studies discussed in the remainder of this paper. Ideally it would be possible to express a complete model within the provided template. The overall goal, however, is to facilitate communicating important modeling assumptions so if the model is too complex to be presented clearly in the template than only the parts most relevant for its main assumptions should be included.

\section{Revisiting the case study}
\label{section:case_study_revisited}
The proposed model class can be used as a tool for interpreting existing models. In this section, we return to GBD and B3 models of U5MR and show how they can be interpreted within the TMMPs model class. The model descriptions are summarized in the TMMP template in Table~\ref{tab-u5mr-comparison}. 

\subsection{Global Burden of Disease (GBD) Model}
As we saw in Section~\ref{section:case_study}, the GBD model is presented as using a three-stage modeling approach in which data go through two smoothing steps before being used in a spatiotemporal Gaussian Process regression model. 

\paragraph{Process Model} To cast the GBD model within our model class, we will start with their third modeling stage, the spatiotemporal Gaussian Process regression model, which can be thought of as a process model. The GBD process model can be understood as having a covariance component, fixed offset, and a Gaussian Process smoothing model component. Define $\eta_{c,t}$ to be the true crisis-free U5MR in country $c$ at time $t$. Then the process model is given by:
\begin{align*}
    \log_{10}\left(\eta_{c,t}\right) = g_2(\bm{X}_{c,t}, \bm{\beta}_c) + a_{c,t} + \epsilon_{c,t}.
\end{align*}
The covariate component is a non-linear function of lag-distributed income per capita (LDI), mean years of education for women of reproductive age (15-49 years), and HIV death rate in ages 0-4 as covariates:
\begin{align}
   g_2(\bm{X}_{c,t}, \bm{\beta}_c) = \exp\left[ \beta_{c,1} \cdot \log(X^{LDI}_{c, t}) + \beta_{2,c} \cdot X^{EDU}_{c, t} + \beta_{3,c} \right] + \beta_{4,c} x^{HIV}_{c, t}.
   \label{eqn:gbd_covariates}
\end{align}
The offset $a_{c,t}$ adjusts the values from the covariate component based on the results of a separate modeling step. The offset values were estimated separately by smoothing the residuals of a mixed-effect model (the second modeling stage, in the original description).

Finally, the smoothing model uses a Gaussian Process prior, with no transformation ($\bm{B} = \bm{I}$, so $\bm{\epsilon}_c = \bm{\delta}_c$)
\begin{align*}
    \bm{\delta}_c \mid \bm{\Sigma}_c \sim N\left(\bm{0}, \bm{\Sigma}_c\right),
\end{align*}
where $\bm{\Sigma}_c$ is given by the Matérn covariance function. 

\paragraph{Parameter Estimation}
Fixed parameter values for the covariate and systematic components come from the first two stages of the GBD modeling approach. The regression coefficients in the covariate component are fixed to values estimated in a separate non-linear mixed effects model (the first modeling stage of the original description of the model.) The offsets $a_{c,t}$ are derived by smoothing the residuals from this model. These residuals are smoothed in time and between countries in the same region. As such, the offsets are estimated using a sort of hierarchical structure, in that information is shared between countries in the same region. The final offsets are calculated as a weighted combination of the smoothed residuals and a local linear fit.

The remaining hyperparameters for the data model and the smoothing component of the process model are largely set to fixed values. For example, the parameters of the Matérn  covariance kernel were fixed depending on a measure of data availability in each country in order that areas of low data availability had higher smoothing than areas with more data. 

\subsection{UN-IGME model}
\label{section:b3-revisited}


\paragraph{Process Model}
Define $\eta_{c,t}$ as the true crisis-free U5MR in country $c$ at time $t$. The process model includes systematic and smoothing components:
\begin{align*}
    \log\left(\eta_{c,t}\right) = g_3(t, \eta_{c,s\neq t}, \bm{\alpha}_c) + \epsilon_{c,t}.
\end{align*}
The systematic component comprises a country-specific slope and intercept:
\begin{align*}
    g_3(t, \eta_{c,s \neq t}, \bm{\alpha}_c) = \alpha_{c,0} + \alpha_{c,1}(t - t_c^*),
\end{align*}
where $t_c^*$ refers approximately to the midpoint of the observation period. 
The smoothing term is defined as
\begin{align*}
    \bm{\epsilon_c} = \bm{B}_c \bm{\delta}_c,
\end{align*}
where $\bm{B}_c$ contains $K_c$ cubic B-spline basis functions placed evenly over the observation period of each country (that is, $B_{c,t,k} = b_{c,k}(t)$ where $b_{c,k}(t)$ is the $k$-th spline basis function for country $c$ evaluated at time $t$). The spline coefficients $ \delta_{c,k}$ for $k=1,2,\hdots, K_c$ follow a RW(2) process. As such, after two levels of differencing, the coefficients are normally distributed with mean zero:
\begin{align}
    \gamma_{c,k} = \triangle_2 \delta_{c,k} \mid \sigma_{\delta, c}^2 \sim N(0, \sigma_{\delta, c}^2)
\end{align}
with
\begin{align}
    \delta_{c,k^*} &= 0, \\
    \sum_{k=2}^{K_c} \triangle_1 \delta_{c,k} &= 0,
\end{align}
with $k_{c^*}$ referring approximately to the spline centered around $t_c^*$, such that $\mathcal{K}_{0,c} = \left\{ k^* \right\}$ and $\mathcal{K}_{1, c} = \left\{2, \dots, K_c \right\}$. These constraints imply that $\bm{\delta}_c$ can be recovered from the $K_c - 2$ second-order deviations $\bm{\gamma}_c$ by
\begin{align}
    \bm{\delta}_c = \left[ \bm{D}^\prime_c (\bm{D}_c \bm{D}^\prime_c)^{-1} \right] \bm{\gamma}_c,
\end{align}
where $\bm{D}$ is a $K_c \times (K_c - 2)$ second order differencing matrix.
%

\paragraph{Projections}
The estimates of the B3 model cover the period of observed data in each country. Projections after the last observed data point in each country are generated by projecting the second order differences of the process model formula. For this model, the second order differences are given by
\begin{align*}
    (\log(\eta_{c,t+1}) - \log(\eta_{c,t})) - (\log(\eta_{c,t}) - \log(\eta_{c,t-1})) &= (\epsilon_{c,t+1} - \epsilon_{c,t}) - (\epsilon_{c,t} - \epsilon_{c,t-1}),
\end{align*}
where the terms $\epsilon_{c,t}$ are obtained by extending the smoothing model to incorporate the projection period:
\begin{align*}
    \epsilon_{c,t} &= \sum_{k=1}^{K_c + P_c} b_{c,k}(t) \delta_{c,k},
\end{align*}
where $b_{c,k}(t)$ is the $k$-th spline function in country $c$ evaluated at time $t$ and $P_c$ refers to the number of spline functions added. The coefficients $\delta_{c, k}$ are extrapolated as follows: 
\begin{align*}
\delta_{c,k + 1} &= 2 \delta_{c,k} - \delta_{c, k -1} + \gamma_{c, k}, \\
\gamma_{c,k} |\Gamma_{c,k},\Theta_{c,k} &\sim N\left(\Gamma_{c, k}, \Theta_{c, k} \right), 
\end{align*}
where
\begin{align*}
\Gamma_{c,k} &= W\cdot G+ (1-W)\cdot \triangle_2\delta_{c,k-1},\\
\Theta_{c,k} &= W\cdot V + (1-W)\cdot \Theta_{c,k-1},
\end{align*}
with $G$ and $V$ equal to the median and variance of the estimates of past 2nd-order differences $\triangle_2 \hat{\delta}_{1:C,1:K_c}$'s respectively, and $\Theta_{c,K_c-1} = \sigma_{\tau, c}^2$.

\paragraph{Parameter Estimation}
The variances $\sigma_{\delta, c}^2$ of the random walk deviations $\delta_{c,k}$ are estimated hierarchically so that information on the degree of smoothing can be shared between countries. Conversely, the parameters $\alpha_{c,0}$ and $\alpha_{c,1}$ controlling the intercept and slope of the systematic component are estimated independently with vague priors. 

\subsection{Comparing the U5MR models}
With both models expressed using the TMMPs notation it is easier to list the main assumptions made in a standardized way and compare assumptions across models. Table \ref{tab-u5mr-comparison} summarizes the process model and parameter estimation strategies of each model. Comparing non-smoothing components, the GBD model uses several covariates (LDI, EDU, and HIV), and an offset estimated in a separate regression, as compared to the use of a linear systematic trend that is non-zero during the observation period in the B3 model. The two models use different smoothing approaches, with the GBD model using a Gaussian process and the B3 model using a second order random walk on B-spline coefficients. Finally, the GBD model fixes model parameters in earlier steps in its estimation procedure, while B3 uses full Bayesian inference to estimate parameter uncertainty during the observation period. B3 uses hierarchical distributions to share information between countries on the variability of the smoothing term only. For the GBD model, the offsets estimated in a separate step by smoothing residuals serve as a way of sharing information across countries.

Using this comparison we can interpret how the models will behave in forward projections. The GBD model's projections are informed by its covariate model and systematic offsets. Given its usage of a stationary smoothing model, deviations away from these components in recent years will converge back to zero, hence the estimates of U5MR will convergence back to the covariate-plus-offset-based estimates. The B3 model short-term projections, on the other hand, derive entirely from the smoothing component, extrapolating  recently observed linear trends (in log space) into the future, while - through the pooling component - in B3 longer term projections, extreme trends converge to a global distribution of past observed rates of change.

\begin{table}[hbtp]
\begin{tabular}{
    | p{0.30\linewidth}
    | p{0.25\linewidth}
    | p{0.45\linewidth} |
} \hline
    & \textbf{GBD} & \textbf{B3} \\
    
    \hline
    
    $\eta_{c,t}$ & 
      crisis-free U5MR &
      crisis-free U5MR \\
      
    \hline
    
    $g_1(\cdot)$ & $\log_{10}$ & $\log$ \\
    
    \hline
    
    Process model formula &
        $g_1(\eta_{c,t}) = g_2(\bm{X}_{c,t}, \bm{\beta}_c) + a_{c,t} + \epsilon_{c,t}$ &
        $g_1(\eta_{c,t}) = g_3(t, \bm{\alpha}_c) + \epsilon_{c,t}$ \\
    
    \hline
    
    \multicolumn{3}{|l|}{\textbf{Covariate Component}} \\
    
    \hline
    
    $g_2(\cdot)$ & non-linear regression formula (Equation \ref{eqn:gbd_covariates}) & $\cdot$ \\
    
    \hline
    
    Covariates & LDI, EDU, HIV & $\cdot$ \\
    
    \hline
    
    \multicolumn{3}{|l|}{\textbf{Systematic Component}} \\
    
    \hline
    
    $g_3(\cdot)$ & $\cdot$ & $\alpha_{c,0} + \alpha_{c,1}(t - t_c^*)$, with $t_c^* \approx$ middle of observation period \\
    
    \hline
    
    $\bm{\alpha}_{c}$ & $\cdot$ & intercept $ \alpha_{c,0}$ and slope $\alpha_{c,1}$ \\
    
    \hline

    \multicolumn{3}{|l|}{\textbf{Offsets}} \\
    
    \hline
    
    $a_{c,t}$ &  offsets obtained from smoothed residuals of a mixed-effects regression model fit & $\cdot$ \\
    
    \hline
    
    \multicolumn{3}{|l|}{\textbf{Smoothing Component} $ \bm{\epsilon_c} = \bm{B}_c \bm{\delta}_c$} \\
    
    \hline
    
    $\bm{B}$ & $\bm{B} = \bm{I}$ & $B_{c,k}=$ cubic B-splines, knots every 2.5 years \\
    
    \hline
    
    $s(t_1, t_2)$ & Matérn & indep. $s(t_1, t_2) = \sigma_{\tau,c}^2 1(t_1 = t_2)$
 \\
    \hline
    
    $r$ & 0 & 2 \\
    
    \hline
    
    $\mathcal{K}_{d,c}$ & $\cdot$ & $\mathcal{K}_{0,c} = \left\{ k^* \right\}$, $\mathcal{K}_{1,c} = \left\{ 2, \cdots, K_c \right\}$ \\
    
    \hline
    
      \multicolumn{3}{|l|}{\textbf{Projections} (if not defaulting to estimation model)} \\
    
    \hline
    
    Projections & $\cdot$ & logarithmic pooling approach: for projections, \begin{eqnarray*}
\triangle_2 \delta_{c,k} &\sim & N\left(\Gamma_{c,k}, \Theta_{c,k}\right), \\
\Gamma_{c,k} &=&W\cdot G+ (1-W)\cdot \triangle_2\delta_{c,k-1},\\
\Theta_{c,k} &=& W\cdot V + (1-W)\cdot \Theta_{c,k-1}.
\end{eqnarray*} \\
    
    \hline
    
    \multicolumn{3}{|l|}{\textbf{Parameter Estimation}} \\
    
    \hline
    
    Fixed & Matérn covariance hyperparameters; $\bm{\beta}_c$ in covariate component & For projections, $G$ and $V$ equal to the median and variance of the estimates of past 2nd-order differences $\hat{\triangle}_2 \delta_{c,k}$'s respectively, $W$ fixed through validation exercise\\
    
    \hline
    
    Vague Priors & $\cdot$ & systematic parameters $\alpha_{c,0}$, $\alpha_{c,1}$ \\
    
    \hline
    
    Informative prior & $\cdot$ & $\cdot$ \\
    
    \hline
    
    Hierarchical distribution & $\cdot$ & smoothing parameters $\sigma^2_{\tau,c}$ \\
    
    \hline
    
    \;\; Distribution $\pi$ & $\cdot$ & normal \\
    
    \hline
    
    \;\; Number of levels in hierarchy & $\cdot$ & 1 \\
    
    \hline
    
    \;\; Hierarchical groupings & $\cdot$ & countries within world \\
    
    \hline 
    
\end{tabular}
\caption{Comparison of the U5MR process model and estimation strategies in the IHME GBD model and the UN-IGME B3 model.}\label{tab-u5mr-comparison}
\end{table}

\section{Additional examples}
\label{section:examples}
In this section we show how models for four different health indicators can be described within our proposed model class.

\subsection{Family Planning Estimation Model}
\label{section:fpem-example}
The Family Planning Estimation Model (FPEM) is a country-level model of contraceptive use rates among married women of reproductive age from 1990-2020 \citep{cahill_modern_2018}, and is an updated version of an earlier model \citep{alkema_national_2013}. The full model breaks down the total use rate into traditional and modern contraceptive method users, and models in addition the unmet need for contraceptives.  For the purposes of this paper we will only examine how it models the total contraceptive use rate, and refer readers to the paper for additional details. 

\begin{figure}[htbp]
    \centering
        \includegraphics[width=0.4\textwidth]{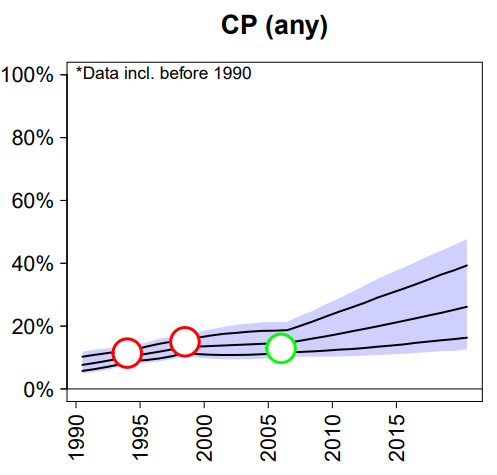}
        \caption{Family Planning Estimation Model (FPEM) estimates for Côte d'Ivoire. The systematic component of the model informs forward projections and estimates in countries with limited data available. Copyright CC BY 4.0.}
    \label{fig:fpem_systematic}
\end{figure}

\paragraph{Process Model} The FPEM process model describes the true trend of total contraceptive use over time in each country. The process model includes a a systematic and smoother component:
\begin{align*}
   \text{logit}(\eta_{c,t}) = g_3(t, \eta_{c,s \neq t}, \bm{\alpha}_c) + \epsilon_{c,t}.
\end{align*}
Adoption of contraception at the country level is expected to start slowly, speed up, then slow down before reaching an asymptote. The model incorporates this expectation of an S-shaped trend through the systematic component. The systematic component encodes the rate of change of a logistic curve. A reference year $t^*$ is fixed, and the systematic trend is propagated forward and backward from the reference year. At $t^*$, we set $g_3(t^*, \eta_{c, s\neq t}, \bm{\alpha}_c) = \Omega_c$, where $\Omega_c$ is the level in the reference year. When $t > t^*$, the systematic component propagates forward:

\begin{align*}
    g_3(t, \eta_{c,s \neq t}, \bm{\alpha}_c) \mid t > t^* &=     \mathrm{logit}\left(\eta_{c,t-1}\right) + \iota_{c,t},\\
    &= 
     \begin{cases}
        \mathrm{logit}\left( \tilde{P}_c \cdot \mathrm{logit}^{-1}\left( \mathrm{logit}\left(\frac{\eta_{c, t-1}}{\tilde{P}_c}  \right) + \omega_c \right) \right), & \text{ when } \eta_{c,t-1} < \tilde{P}_c \\
       \mathrm{logit}\left(\eta_{c,t-1}\right), & \text{otherwise,}
    \end{cases} 
\end{align*}
where $\tilde{P}_c$ is an asymptote, $\omega_c$ a rate parameter, and $\bm{\alpha}_c = \left\{\tilde{P}, \omega_c, \Omega_c \right\}$. A similar equation is applied when $t<t^*$ to extend the systematic trend backwards from the reference year. This component encodes the assumption that contraceptive adoption in each country follows the same shape, but may differ in its timing, rate, peak adoption. The effect of the systematic component can be seen in the model's projections and fits in data-sparse countries. In Côte d'Ivoire, the model projects an increase in contraceptive use because the country's systematic trend, with hyperparameters informed by other countries in the region, puts the country on the verge of its contraceptive transition (Figure \ref{fig:fpem_systematic}).

Deviations from the expected rate of change given by the systematic component are captured via an AR(1) smoothing component.  Using the notation of our proposed model class, this can be expressed as placing a multivariate normal distribution on the vector of deviations:
\begin{align*}
    \bm{\epsilon}_c \mid \bm{\Sigma}_c \sim N(\bm{0}, \bm{\Sigma}_c),
\end{align*}
where $\bm{\Sigma}_c$ is given by the AR(1) autocovariance function. These deviations are added to the systematic component, which for time $t > t^*$ depends on $\eta_{c,t-1}$. As such, the smoothing terms $\epsilon_{c,t}$ represent deviations in the rate of change of modern contraceptive use within a country. Because these smoothing terms accumulate over time, variance increases with the projection horizon.


\paragraph{Parameter Estimation}
FPEM introduces many country-level parameters that need to be estimated. The systematic component for FPEM has parameters for the asymptote, rate, and timing of each country's logistic curve that describes its adoption of contraceptives. Hierarchical distributions share information about each parameter between countries. The asymptote parameters have one level of hierarchy:
\begin{align*}
    \tilde{P}_c \mid \tilde{P}_w, \sigma_{\tilde{P}}^2 &\sim N\left(\tilde{P}_w, \sigma^2_{\tilde{P}}\right),
\end{align*}
where $\tilde{P}_w$ is the world mean asymptote, and $\sigma^2_{\tilde{P}}$ describes the variation around that mean. The rate parameters have three levels of hierarchy, in that they are nested first within sub-regions and regions. The hierarchical model for the rate parameter $\omega_c$ is given by:
\begin{align*}
    \omega_c \mid \omega_{s[c]}, \sigma^2_{\omega_s} &\sim N(\omega_{s[c]}, \sigma^2_{s}), \\
    \omega_s \mid \omega_{r[c]}, \sigma^2_{\omega_r} &\sim N(\omega_{r[c]}, \sigma^2_{r}), \\
    \omega_r \mid \omega_w, \sigma^2_{\omega_w} &\sim N(\omega_w, \omega^2_w),
\end{align*}
where $s[c]$ indexes the sub-region and $r[c]$ indexes the region of country $c$.  The hierarchical structure for the timing parameters $\Omega_c$ depends on whether the country is classified as developing or developed. For developing countries, three levels of hierarchy are used (sub-region, region, and world); for developed countries, only one level is used. The difference in hierarchical structure between $\tilde{P}$ and $\Omega_c$, $\omega_c$ arises from substantive knowledge, in that rate is expected to differ regionally, and the timing may vary for developing countries, while the asymptote is expected to be less variable. 

\subsection{Neonatal mortality rate model}
\label{section:example-nmr}
\cite{alexander_global_2018} model the neonatal mortality rate in 195 countries from at latest 1990 to 2015, and was adopted by the United Nations Inter-agency Group for Child Mortality Estimation. The model setup incorporates the strong relationship between NMR and U5MR: as U5MR increases, the ratio of neonatal to other child mortality tends to decrease. As such, this model provides an example of how covariates can be used in our proposed model class.

\paragraph{Process Model} Rather than model the NMR directly, the process model specifies the ratio of the neonatal mortality rate to the rate of non-neonatal deaths. This was motivated by the strong relationship between the ratio and U5MR, and also to constrain the NMR to be smaller than the U5MR. 

\begin{figure}[htbp]
    \centering
        \includegraphics[width=0.5\textwidth]{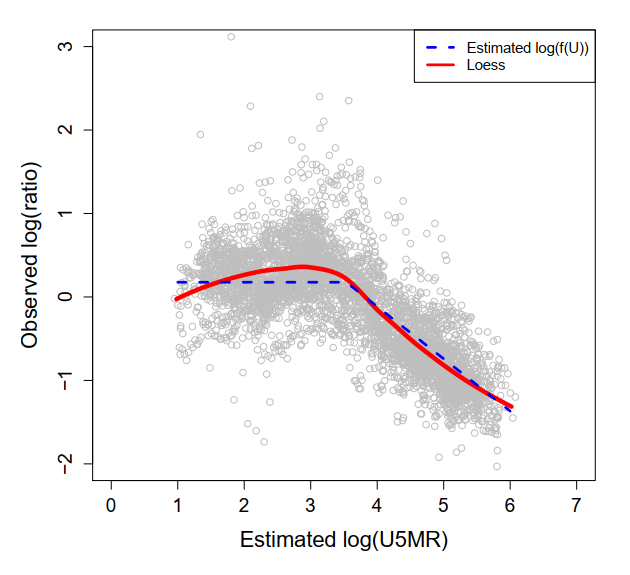}
        \caption{The relationship between $\log(\mathrm{U5MR})$ and the ratio of neonatal deaths to non-neonatal deaths observed in the dataset used for the neonatal mortality rate model of Alexander et al. (2018). Copyright CC BY 3.0 DE.}
    \label{fig:alexander_2018_nmr_eda}
\end{figure}

We define $\eta_{c,t} = N_{c,t} \slash U_{c,t}$, where $N_{c,t}$ and $U_{c,t}$ are the neonatal and non-neonatal death rates (under age 5) in country $c$ at time $t$. The process model includes a covariate 
and smoothing component:
\begin{align*}
    \log(\eta_{c,t}) = g_2(X_{c,t}, \bm{\beta}_c) 
    + \epsilon_{c,t},
\end{align*}
where $X_{c,t}$ is the U5MR for country $c$ at time $t$. The authors found through exploratory data analyses that the relationship between U5MR and the log ratio is constant up to a cutoff U5MR value and then decreases linearly (Figure \ref{fig:alexander_2018_nmr_eda}). As such, they chose a piecewise linear form for the function $h$:
\begin{align*}
    g_2(X_{c,t}, \bm{\beta}_c) = \beta_{c,0} + \beta_1 \cdot \left( \log(X_{c,t}) - \log(\beta_2) \right)\bm{1}_{[X_{c,t} > \beta_2]},
\end{align*}
where $\beta_{c,0}$ and $\beta_1$ are intercept and slope parameters, and $\beta_2$ is the cutoff U5MR value at which point the relationship becomes constant. 

The deviations $\eta_{c,t}$ capture trends in the data not explained by the covariate component and are modeled with B-splines, similar to the B3 model. 
In TMMP class notation we can write
\begin{align*}
    \bm{\epsilon}_c = \bm{B}_c \bm{\delta}_c,
\end{align*}
where $\bm{B}_c$ contains $K_c$ cubic B-spline basis functions, placed at evenly spaced knot locations. Sufficient knots are placed in each country to cover the period for which data are available. The spline coefficients $\bm{\delta}_c$ are modeled with an RW(1) process, which after one level of differencing is multivariate normally distributed with mean zero. That is,
\begin{align}
    \gamma_{c,k} = \triangle \delta_{c,k} \mid \sigma^2_{\gamma, c} \sim N(0, \sigma_{\gamma, c}^2),
\end{align}
with 
\begin{align}
    \sum_{k=1}^{K_c} \delta_{c,k} = 0,
\end{align}
such that $\mathcal{K}_{0, c} = \left\{ 1, \dots, K_c\right\}$. Note that the sum-to-zero constraint implies that $\bm{\delta}_c$ can be recovered from the $K_c-1$ deviations $\bm{\gamma}_c$ by
\begin{align}
    \bm{\delta}_c = \left[ \bm{D}^\prime_c \left( \bm{D}_c \bm{D}^\prime_c \right)^{-1} \right] \bm{\gamma}_c,
\end{align}
where $\bm{D}$ is a $K_c \times (K_c - 1)$ first-order differencing matrix. 



\begin{figure}[htbp]
    \centering
        \includegraphics[width=0.7\textwidth, trim=0 100 0 0, clip]{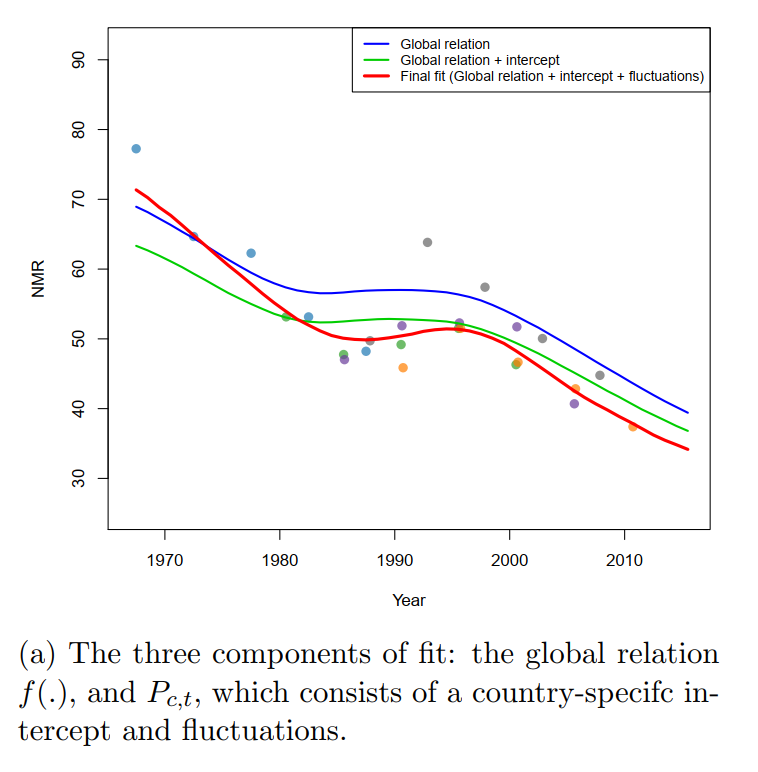}   
        \caption{Example estimates for neonatal mortality from Alexander and Alkema (2018). Global relation (blue) shows the estimates from the covariate component of the process model, global relation + intercept (green) the covariate estimates plus a country-specific intercept, and final fit (red) shows the final estimates that include the smoothing model. The smoothing model has the most impact within the data support, as it modifies the expected trend based on covariates to better fit the observed data. Copyright: CC BY 3.0 DE.}
    \label{fig:alexander_2018_nmr_process_model}
\end{figure}

Figure \ref{fig:alexander_2018_nmr_process_model} shows model estimates separated into covariate and smoothing components, which illustrates how projections are influenced by the relationship between U5MR and covariates. The projection follows the expected trend based on covariates closely, with a country-specific offset. The smoothing component has a larger effect on the model fit where there are data available, showing how the model modifies the expected trend from covariates to fit the observed NMR data.

\paragraph{Projections}
The main model produces estimates that cover the period of observed data in each country. As per the TMMP specification, for the NMR model with the non-stationary RW(1) smoother with $r=1$, projections for the period after the last observed data point in each country to 2015 are obtained by projecting first order differences that follow from the process model expression. Here the first order differences are given by
\begin{align}
    \eta_{c,t+1} - \eta_{c,t} &= \beta_1 \cdot \left( \log(X_{c,t+1})\bm{1}_{[X_{c,t+1} > \beta_3]} - \log(X_{c,t}) \bm{1}_{[X_{c,t} > \beta_3]}\right) + \epsilon_{c,t+1} - \epsilon_{c,t},
\end{align}
where the smoothing terms $\epsilon_{c,t}$ are obtained from extending the spline smoothing model 
\begin{align}
    \epsilon_{c,t} &= \sum_{k=1}^{K_c+P_c} b_{c,k}(t) \delta_{c,k},
\end{align}
where $P_c$ refers to the total number of spline functions added and the $\delta_{c,k}$ are projected based on the RW(1) process:
\begin{align}
\delta_{c,k} &= \delta_{c,k-1} + \gamma_{c,k},\\
\gamma_{c,k}|\sigma^2_{\gamma,c} &\sim N(0, \sigma^2_{\gamma,c}). 
\end{align}


                                                                                                    
\paragraph{Parameter Estimation}
The smoothing model estimates some country-level parameters independently and others hierarchically, which encodes assumptions about how under-5 mortality rates are related across countries. The intercepts $\beta_{c, 0}$ were estimated hierarchically, as were the variances of the smoothing model deviations.
All other hyperparameters were assigned vague priors.

\subsection{Bayesian maternal mortality model}

 The Bayesian Maternal Mortality Model, referred to as `Bmat', is used to estimate the maternal mortality ratio (MMR, maternal deaths per 100,000 live births) in countries from 1985 to 2017 \citepalias{alkema_bayesian_2017, mmeig2015, mmeig2019}. The model is an extension of a previous multilevel regression modeling approach \citep{wilmoth2012new} used by the United Nations Maternal Mortality Estimation Inter-agency group (UN MMEIG). Bmat incorporates covariates similarly to the earlier model, while also allowing data-driven deviations from the expected covariate trend. The Bmat model handles AIDS and non-AIDS maternal deaths separately; for the purposes of this paper we will focus on the model of non-AIDS maternal deaths.

\paragraph{Process Model} Let $\eta_{c,t}$ represent the proportion of non-AIDS deaths to women of reproductive age that are of a maternal cause in country $c$ at time $t$. Bmat models the expected trend in $\eta$ by a  combination of covariate and smoothing model components:
\begin{align*}
    \log\left(\eta_{c,t}\right) = g_2(\bm{X}_{c,t}, \bm{\beta}_c) + \epsilon_{c,t},
\end{align*}
where $\bm{X}_{c,t}$ is a set of covariates for country $c$ at time $t$, and $\bm{\beta}_c$ are associated regression coefficients.

The covariate model is based on the earlier UN MMEIG regression model, and is given by
\begin{align*}
    g_2(\bm{X}_{c,t}, \bm{\beta}_c) = \beta_{c,0} + \beta_{1} \log(X^{GDP}_{c,t}) + \beta_{2} \log(X^{GFR}_{c,t}) + \beta_{3} X^{SAB}_{c,t}.
\end{align*}

The deviations are smoothed with an ARIMA(1, 1, 1) process, which is a stationary autoregressive moving average (ARMA) process after differencing once ($r = 1$). In TMMP notation
\begin{align*}
	\bm{\epsilon}_c &= \bm{\delta}_c,\\
	\delta_{c,t}    &= 0, \text {for } t = 1990,\\
	\triangle_1  \bm{\delta}_c \mid \bm{\Sigma}_c &\sim N(\bm{0}, \bm{\Sigma}_c),
\end{align*}
where  the covariance matrix $\bm{\Sigma}_c$ is specified via an autocovariance function $s$ that captures the autocovariance of an ARMA(1,1) process, parametrized with country-specific stationary variance $\sigma_{\delta, c}^2$, autoregressive parameter $0 \leq \rho \leq 1$ and moving average parameter $-1 \leq \theta \leq 0$.

\begin{figure}[htbp]
    \centering
        \includegraphics[width=0.8\textwidth]{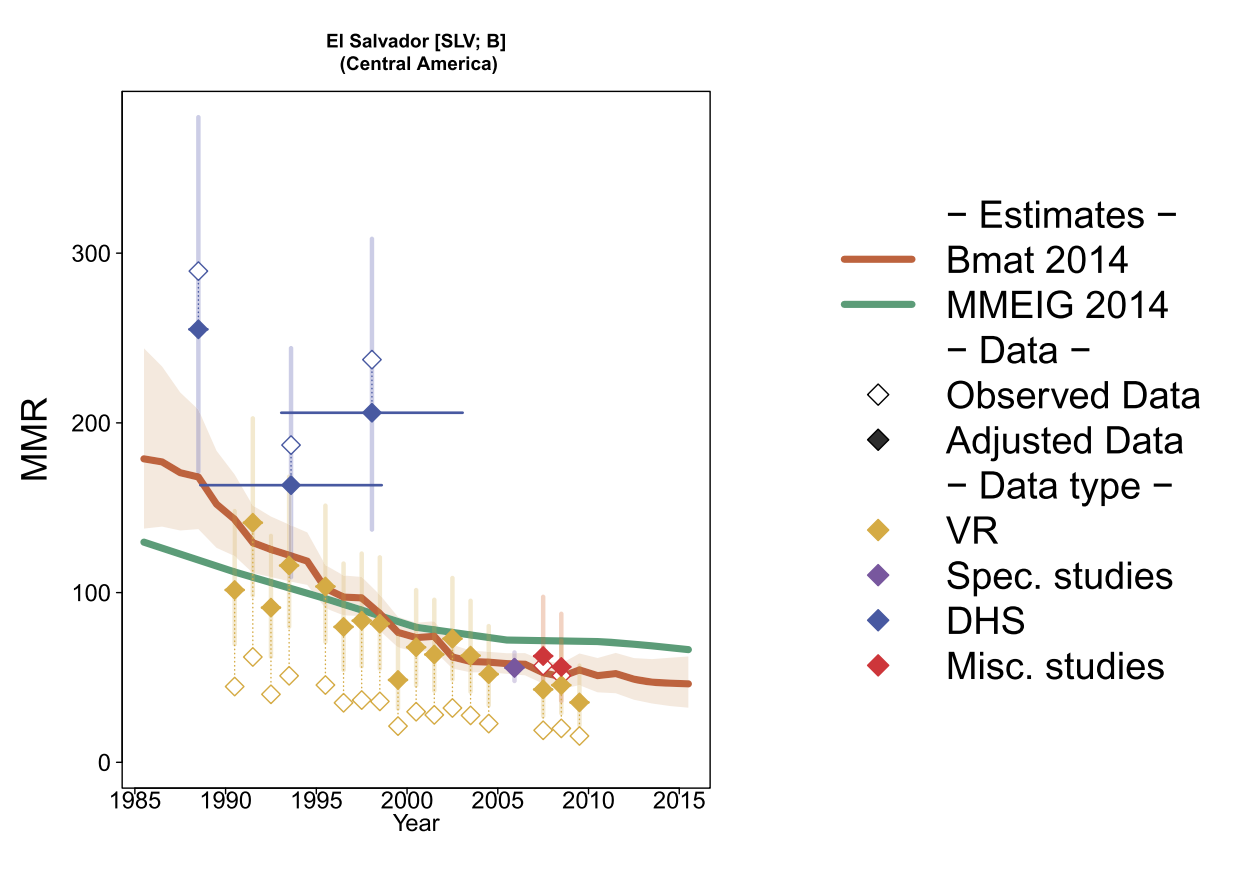}
        \caption{Bmat estimates of maternal mortality for El Salvador. Bmat estimates are displayed in red and follow bias-adjusted data more closely than covariate-driven estimates based on the UN MMEIG 2014 model. 
        Reproduced with permission.}
    \label{fig:alkema_2017_mmr_el_salvador}
\end{figure}

The addition of a smoothing model increases the flexibility of the process model. 
Model estimates combine covariate trends and patterns in the observed maternal mortality data, as illustrated in Figure~\ref{fig:alkema_2017_mmr_el_salvador} for El Salvador: the model-based estimates deviate from the expected covariate-based trend and better fits historical DHS data and high-quality VR data, when available.

%
%
\paragraph{Parameter Estimation} 
The intercepts $\beta_{c,0}$ are estimated hierarchically with two levels (country within region within world). The other regression coefficients are shared between all countries and are given a vague prior. The country-specific variance parameter of the ARIMA(1, 1) smoothing model is estimated hierarchically in order to share information between countries:
\begin{align*}
    \sigma_{\delta, c} = \sigma_{\delta, w} \cdot (1 + \lambda_c) \\
    \lambda_c \mid \sigma_\lambda^2 \sim TN_{(-1, 2)}(0, \sigma_\lambda^2).
\end{align*}
The parameter $\sigma_{\delta, w}$ represents a central value of the variances across all countries, and $\lambda_c$ is a country-specific multiplier. A truncated normal prior is placed on $\lambda_c$, limiting its values to fall between $-1$ and $2$. This hierarchical structure allows different countries to have differing levels of smoothness in their fluctuations, while still sharing information between countries. 

\subsection{Bayesian model for subnational mortality}

\cite{alexander_flexible_2017} developed a model for estimating age-specific mortality rates at the subnational level, focusing on obtaining estimates by county in the United States. Unlike the other examples mentioned in this paper, this model set-up is more of a 'traditional' demographic model as it deals explicitly with modeling age patterns in mortality over time, rather than focusing on modeling temporal trends in an aggregate indicator of mortality. The challenge with estimating mortality age schedules at the subnational level is that, when populations are small, stochastic variation is high, and so the underlying mortality risk curve may be unclear or uncertain from the observed data. 

In essence, \cite{alexander_flexible_2017} build on the traditional demographic method of using model age schedules of mortality, but place this within a Bayesian framework which allows for stochastic estimates and forecasts and increased flexibility in estimates. We include it as an example here to show that both demographic models and models for global health indicators can be expressed in the same generalized TMMP framework. 

\paragraph{Process Model} For the process model, $\eta_{a,c,t}$ is defined as the mortality rate for age $a$ in county $c$ at time $t$. The process model incorporates a covariate and smoothing component:
\begin{align*}
    \log\left(\eta_{a,c,t}\right) = g_2(\bm{X}_{a}, \bm{\beta}_{c, t}) + \epsilon_{a, c, t},
\end{align*}
where $\bm{X}_{a} = \left\{ X_{a, 1}, X_{a, 2}, X_{a, 3} \right\}$ are the first three principal components of a set of standard mortality curves, and $\bm{\beta}_{c,t}$ are the associated regression coefficients which differ by county and time point. The regression is equation is given by:
\begin{align*}
    g_2\left(\bm{X}_a, \bm{\beta}_{c, t}\right) = \beta_{c, t, 1} \cdot X_{a,1} + \beta_{c, t, 2} \cdot X_{a,2} + \beta_{c, t, 3} \cdot X_{a,3}.
\end{align*}
Whereas previous model examples with covariates have focused on relating trends in country-level indicators, the covariates here capture the main sources of variation in mortality over age, which allows for plausible estimates over age-specific mortality to be imputed, and also 'smooths out' noisy age-specific mortality curves that are commonly observed in small populations. 

The $\epsilon_{a,c,t}$ term was included in the model to account for overdisperson commonly seen in observed mortality rates. Unlike previous examples, no temporal model is placed on the deviations $\epsilon_{a,c,t}$ to smooth the deviations over time. Instead, the deviations are smoothed within age groups:
\begin{align*}
    \epsilon_{a, c, t}|\sigma_a^2 \sim N(0, \sigma_a^2),
\end{align*}
where $\sigma_a^2$ is the variance for age group $a$. 

\begin{figure}[htbp]
    \centering
        \includegraphics[width=0.6\textwidth]{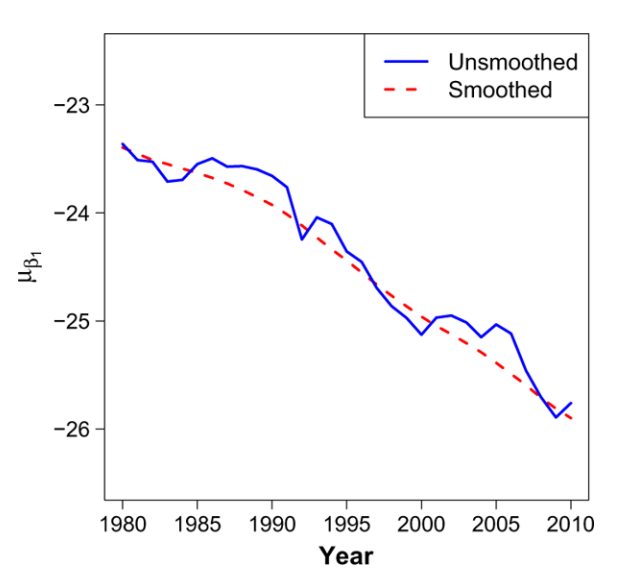}
        \caption{Estimated hyperparameters from \cite{alexander_flexible_2017} with and without temporal smoothing. Reproduced with permission. }
    \label{fig:subnational_mortality}
\end{figure}

\paragraph{Parameter Estimation} The process model requires estimating regression coefficients $\beta_{c,t,p}$ for every principal component, county, and time point. To share information across counties, a hierarchical prior is placed on the regression coefficients:
\begin{align*}
    \beta_{p, c, t}|\mu_{p, t}, \sigma^2_{p, t} \sim N\left(\mu_{p, t}, \sigma^2_{p, t}\right).
\end{align*}
Further structure is placed on the hyperparameters $\mu_{p,t}$ in order to smooth over time, with the assumption being that the underlying expected trend in variation in each dimension exhibits some smooth trend over time. 
\begin{align*}
    \mu_{p, t}|\mu_{p, t - 1}, \mu_{p, t- 2}, \sigma^2_{\mu} \sim N(2\mu_{p, t - 1} - \mu_{p, t- 2}, \sigma^2_{\mu}).
\end{align*}
The model was fitted with and without this temporal smoothing structure are shown in Figure \ref{fig:subnational_mortality}. Non-informative priors were set on the remaining process model hyperparameters.

\section{Discussion}
\label{section:discussion}

In this paper we have introduced a general model class, ``Temporal Models for Multiple Populations" (TMMPs), which encompasses many existing demographic and health models. The key structural feature of the class is that it makes a distinction between the data and process model, separating the expected trend of an indicator from details of how the observed data were generated. Thus, the process model expresses the modeler's assumptions about the dynamics driving the latent value of the indicator, and the data model  describes assumptions about noise and error in the observed data.  Once a model is shown to fall into this model class, it is easier to compare the assumptions made in different models.  

We showed how six existing models of demographic and health indicators fit into the model class. The example models were for a diverse set of outcomes: total contraceptive use, under-5 mortality, maternal mortality, neonatal mortality, and sub-national mortality rates. These models incorporated a variety of modeling approaches, including Gaussian process regression, ARIMA time series, and penalized spline regressions. The Global Burden of Disease Study U5MR model in particular might appear to be the most different of the example models based on how it was originally presented as a three-stage modeling procedure \citep{dicker_global_2018}. However, we showed how this model and the other example models can be expressed under our modeling class, which facilitates documentation of assumptions and across-model comparisons. We recommend that the writing of model assumptions for TMMPs in a standardized way is considered for a future version of GATHER, referring to Guidelines for Accurate and Transparent Health Estimates Reporting \citep{stevens2016guidelines}. We provided a template for doing so in Appendix~B.

The models we included as examples in this paper are by no means an exhaustive list of the models than can be described within our framework. Many spatial models, which directly model spatial correlation between units, can be expressed in our class. Spatial smoothing models following the work of Knorr-Held (\citeyear{knorr2000bayesian}) in particular are an example of a modeling approach that falls into our model class. Additional research is needed to investigate how other modeling approaches relate to TMMPs. The structure of the TMMP model class is flexible and extendable, making it possible to propose new components as necessary to capture models currently outside of the class. 

So far we have focused on describing the model class and showing how existing models fall into it, rather than evaluating the assumptions made by each model. In future work we plan to compare in more detail the effects of the particular choice for each component on the resulting estimates. A better understanding of the effects of various smoothing models, for example, is helpful both for analyzing the behavior of existing models and for developing new models.


The number of models developed for providing demographic and health indicator estimates has perhaps grown faster than tools to interpret their results and how they relate to one another. The model class proposed in this paper is one tool that can be used to understand model assumptions. We hope it will facilitate interpreting, comparing, and contrasting existing models that fall within the model class as well as the development of new modeling approaches.

\bibliography{references}

\end{document}


\title{Supplementary Material}

\section*{Appendix A} 

\subsection*{Properties of stationary smoothing models, with $r=0$}
\label{section:stationary_models}
When there is no differencing in the smoothing model ($r = 0$) then by construction the distribution of the smoothing terms is stationary: the unconditional first and second moments of $\bm{\epsilon}_c$ will not depend on time. As we will see shortly, this implies that in the absence of data the smoothing term will be centered around zero. This is important for understanding the behavior of the model in projections: the smoothing model will contribute to uncertainty in the projections, but since it will eventually revert to mean zero the projections will be centered around the other process model components. As we will see in the next section, this is not true when $r>0$ because the smoothing terms are no longer stationary.

To make our understanding of the behavior of these stationary smoothing models more precise and to facilitate comparison across smoothing models, we can look at their conditional distributions. This helps particularly in exploring the implied projections of each smoothing model given a last observed data point. 

\begin{figure}[htp]
    \centering
        \includegraphics[width=\textwidth]{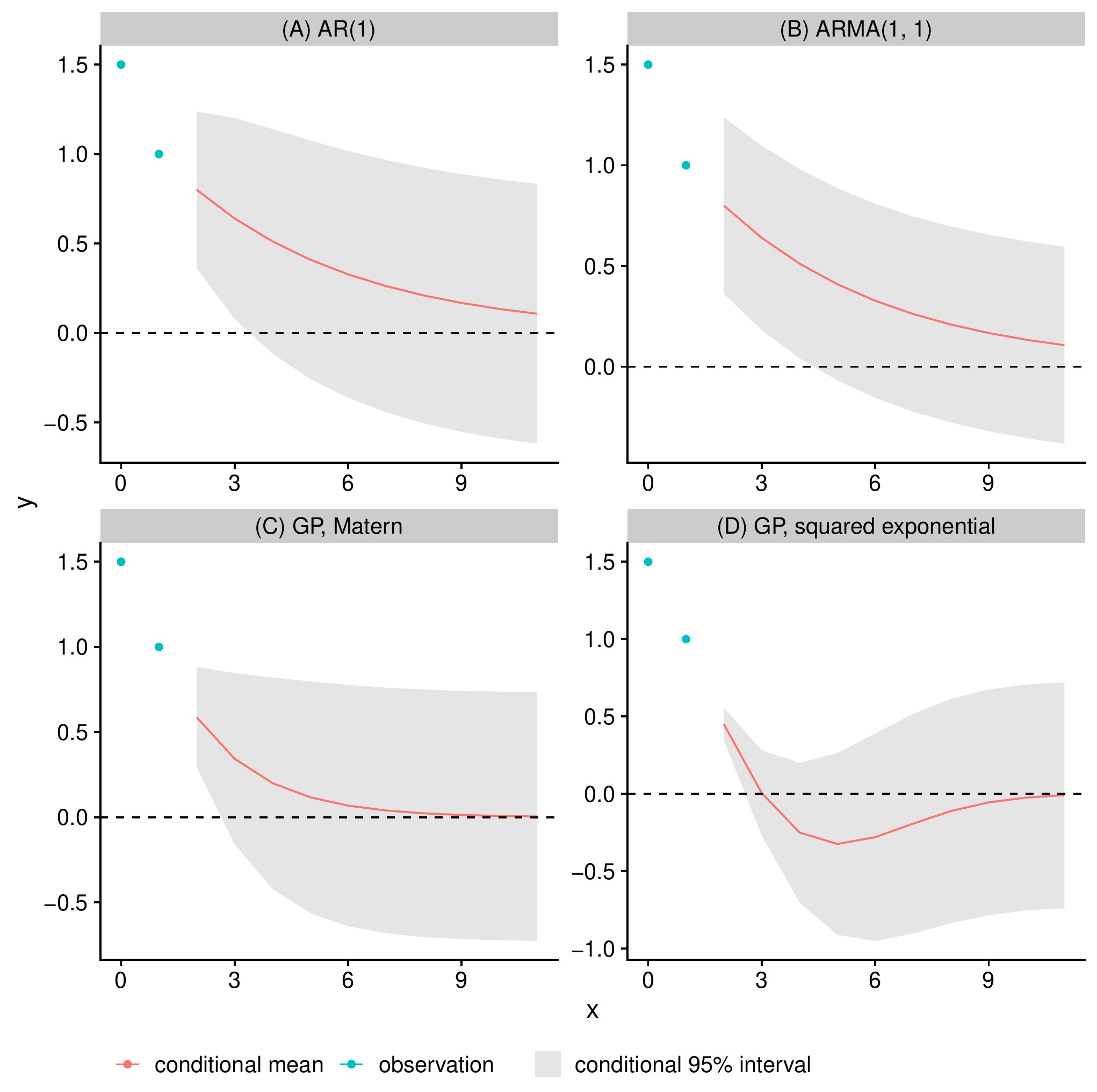}
    \caption{Comparison of the conditional distributions of four smoothing models based on two observed data points. The parameters for the smoothers are: (A) AR(1), $\rho=0.8, \kappa^2=0.05$. (B) ARMA(1, 1), $\rho = 0.8$, moving average parameter $\theta = 0.8$, $\kappa^2=0.05$. (C) GP, Matérn, $\kappa^2 = 0.05 \slash (1 - \rho^2), \rho = 0.8, \nu = 3\slash 2, \ell = 3$. (D) GP, squared exponential, $\kappa^2 = 0.05 \slash ( 1 - \rho^2), \rho = 0.8, \ell = 3$.}
    \label{fig:conditional_dists}
\end{figure}

Suppose we have observed data up to time $t^*$, and let $\bm{\epsilon_{t^*}} = [\epsilon_1, \cdots, \epsilon_{t^*}]$. We set $\bm{B} = \bm{I}$ to simplify the analysis, which means $\bm{\epsilon} = \bm{\delta}$ (the two are interchangeable, and we drop the subscript $c$ for simplicity). Similarly define $\bm{\delta}_{t^*} = [\delta_1, \cdots, \delta_{t^*}]$. We placed the restriction that $\bm{\delta}$ be multivariate normally distributed, which means there is a closed-form solution for the distribution of $\bm{\delta}$ conditional on the observed data. The conditional mean of $\delta_t$ given $\bm{\delta}_{t^*}$ for $t > t^*$ is given by
\begin{align*}
    \mathrm{E}[\delta_t \mid \bm{\delta}_t^*] = L^\top \Sigma^{-1} \bm{\delta}_{t^*},
\end{align*}
where $L$ is the Gram matrix derived from the covariances between  $\delta_t$ and every element of $\bm{\delta}_{t^*}$ ($L_{i} = s(t, i)$ for $i=1,\cdots,t^*$), and $\Sigma$ is the Gram matrix derived from the covariances between every pair in $\bm{\delta}_{t^*}$. We restricted the covariance function of the smoothing model to depend only differences in time, and that it goes to zero as the differences in time grows. This ensures that the conditional mean is guaranteed to converge to zero as $t\rightarrow\infty$. The manner in which the conditional means converge is determined by the specific covariance function. For example, the sparsity of $\Sigma^{-1}$ for an AR(1) kernel makes it straightforward to derive the conditional distribution:
\begin{align*}
    \mathrm{E}[\delta_t \mid \bm{\delta}_{t^*}] &= \delta_{t*} \cdot \rho^{(t - t^*)}.
\end{align*}
In projections, an AR(1) process depends only on the last observed value ($\delta_{t^*})$, converging back to zero as $\rho^{(t - t^*)}$ converges to zero. This is a consequence of the sparsity of the precision matrix for the AR(1) covariance. Covariance functions that yield non-sparse precision matrices do not lead to conditional means with simple forms like the AR(1) process. This leads to more complex behavior: for example, the squared exponential kernel can project trends in the previously observed data before returning to zero. Figure \ref{fig:conditional_dists} compares the conditional behavior of several covariance functions.

\subsection*{Non-stationary smoothing models, with $r>0$}
\label{section:non-stat-smoothers}

When the degree of differencing is greater than zero ($r>0$) then the resulting smoothing models are not stationary (their unconditional first and second moments will depend on time). However, differencing can yield a stationary process. For example, a RW(1) process, with 
\begin{align*}
    \delta_t|\delta_{t-1}, \sigma \sim N(\delta_{t-1}, \sigma^2),
\end{align*}
after one level of differencing becomes
\begin{align*}
    \triangle \delta_t \mid \sigma \sim N(0, \sigma^2).
\end{align*}
Therefore within our framework the RW(1) model can be expressed as $r=1$ (one level of differencing to yield a stationary process) and $\bm{\Sigma}_c = \sigma^2\bm{I}$ (the covariance matrix of the differenced stationary process is diagonal). Similarly, a RW(2) process after two levels of differencing is stationary, so we describe it as $r=2$ and $\bm{\Sigma}_c = \sigma^2\bm{I}$. Autoregressive integrated moving average (ARIMA(p,d,q)) models are obtained by setting $r = d$ and using the respective autocovariance function in $\bm{\Sigma}_c$. For example, the autoregressive integrated ARI(1,1) model is given by $r=1$ and the AR(1) covariance function.

Unconstrained non-stationary smoothing processes can be written according to the TMMP process model specification by a reparametrization of such processes into a zero-constrained process and a structural component. 
For example, an unconstrained RW(1) ($d = 1$) for years $\mathcal{K}_{1,c}$ can be reparametrized into one that includes a sum-to-zero constraint by introducing as parameters the mean of the process and the deviations away from the mean. This parametrization is most clearly expressed using a first order differencing matrix $\bm{D}$ with $D_{i,i} = -1$, $D_{i, i+1} = 1$, and is zero everywhere else. With $\bm{D}$ and  $\mathcal{K}_{1,c} = \{1,2, \hdots, t^*\}$, we can write 
\begin{align*}
     \bm{\delta}_{t^*} = \alpha_0 + [\bm{D}^\prime (\bm{D}\bm{D}^\prime)^{-1}]\bm{\gamma}_{t^*-1},
 \end{align*}
 where $\alpha_0 = 1/t^{*}\sum_t \delta_t $ and $\gamma_t = \delta_t - \delta_{t-1}$ for $t=2, ..., t^*$. Hence, an unconstrained RW(1) process $\delta_t$ can be rewritten as the combination of a smoothing process $\gamma$ with $\sum_{t \in \mathcal{K}_{1,c}} \triangle_d \gamma_{c,t} = 0$ and a systematic component $g_3( t, \alpha_c) = \alpha_0$ for $t \in \mathcal{K}_{d,c}$. Moving the intercept from the smoothing component to the systematic component also helps to clarify the behavior of the model: an overall level during the period associated with $\mathcal{K}_{1,c}$ is estimated as a systematic trend, and deviations are left to the smoothing model. Similarly, an unconstrained RW(2) process can be reparametrized into the mean and rate of change of the process during sets of years $\mathcal{K}_{1,c}$ and $\mathcal{K}_{2,c}$.  
 
Non-stationary smoothing models do not necessarily revert to mean zero in the absence of data, unlike the models we saw in the previous section. As such, non-stationary models can influence the trend of projections. The RW(1) model will extend forward the last observed data point, and the RW(2) model will extend forward a linear trend based on $\triangle \delta_t \mid \sigma, \triangle \delta_{t-1} \sim N(\triangle \delta_{t-1}, \sigma^2)$. The behavior of projections will therefore be determined by the interplay between the smoothing and the covariate and systematic process model components that extend past $\mathcal{K}_{d,c}$.

\section*{Appendix B}\label{section-modelspecsexamples}\label{section-modelspecs}

The following tables are templates for specifying models that fall within the TMMP framework. Four example specifications are provided for each of the additional examples described in the main text.

\begin{longtable}[c]{
    |p{0.25\linewidth}
    |P{0.70\linewidth}|
} \hline
    & \textbf{FPEM} \\
    
    \hline
    
    Citation & \cite{cahill_modern_2018} \\
    
    \hline
    
    $\eta_{c,t}$ & total contraceptive use rate \\
    
    \hline
    
    $g_1(\cdot)$ & $\mathrm{logit}$ \\
    
    \hline
    
    Process model formula &
        $\mathrm{logit}(\eta_{c,t}) = g_3(\cdot) + \epsilon_{c,t}$ \\
    
    \hline
    
     \multicolumn{2}{|l|}{\textbf{Covariate Component}} \\
    
    \hline
    
    $g_2(\cdot)$ & 
        $\cdot$ \\
    
    \hline
    
    Covariates & 
        $\cdot$ \\
    
    \hline
    
    \multicolumn{2}{|l|}{\textbf{Systematic Component}} \\
    
    \hline
    
    $g_3(\cdot)$ &  Logistic curve: \newline 
        $g_3(\cdot) = \Omega_c$ when $t = t^*$, and for $t>t^*$: \newline
        $g_3(\cdot) = \mathrm{logit}\left( \eta_{c,t-1} \right) + \delta_{c,t} $ \\
        $=\begin{cases}
            \mathrm{logit}\left( \tilde{P}_c \cdot \mathrm{logit}^{-1}\left( \mathrm{logit}\left(\frac{\eta_{c, t-1}}{\tilde{P}_c}  \right) + \omega_c \right) \right), & \text{ when } \eta_{c,t-1} < \tilde{P}_c \\
           \mathrm{logit}\left(\eta_{c,t-1}\right), & \text{otherwise,}
        \end{cases}$
    \newline
        where $\bm{\alpha}_c = \left\{ \tilde{P}_c, \omega_c, \Omega_c \right\}$. \\
    
    \hline
    
    $\bm{\alpha}_c$ & 
        $\tilde{P}, \omega_c$, $\Omega_c$ \\
    
    \hline
     
    \multicolumn{2}{|l|}{\textbf{Offsets}} \\
    
    \hline
    
    $a_{c,t}$ &
        $\cdot$ \\
        
    \hline
    
    \multicolumn{2}{|l|}{\textbf{Smoothing Component}} \\
    
    \hline
    
    $\bm{B}$ & 
        $\bm{B} = \bm{I}$ \\
    
    \hline
    
    $s(t_1, t_2)$ & 
        AR(1); $s(t_1, t_2) = \frac{\rho^{|t_1 - t_2|}}{\sigma^2 (1 - \rho^2)}$ \\
    
    \hline
    
    $r$ & 0 \\
    
    \hline
    
    $\mathcal{K}_{d,c}$ & 
        $\cdot$ \\
    \hline
    
    \multicolumn{2}{|l|}{\textbf{Parameter Estimation}} \\
    
    \hline
    
    Fixed & 
        $\cdot$ \\
    
    \hline
    
    Vague Priors & 
        $\cdot$ \\
    
    \hline
    
    Informative Priors & 
        $\cdot$ \\
    
    \hline
    
    Hierarchical model & 
        systematic parameters $\tilde{P}, \omega_c, \Omega_c$ \\
        
     \hline
    
    Hierarchical distribution $\pi$ & 
        normal \\
    
    \hline
    
    Number of levels in hierarchy & 
        $\tilde{P}$: 1 \newline 
        $\omega_c$: 3 \newline
        $\Omega_c$, developing countries: 3 \newline
        $\Omega_c$, developed countries: 1 \\
    
    \hline
    
    Hierarchical groupings & 
        $\tilde{P}$: countries within world  \newline 
        $\omega_c$: countries within sub-region, region world \newline $\Omega_c$, developing countries: countries within sub-region, region, world \newline
        $\Omega_c$, developed countries: countries within world \\
        
    \hline
    
    \multicolumn{2}{|l|}{\textbf{Projections}} \\
    
    \hline
    
    Projections & 
    $\cdot$ \\
    
    \hline
\caption{TMMP specification for the Family Planning Estimation Model \citep{cahill_modern_2018}.}
\label{tab-example-comparison}
\end{longtable}

\begin{longtable}[c]{
    |p{0.25\linewidth}
    |P{0.7\linewidth} |
} \hline
    & \textbf{NMR} \\
    
    \hline
    
    Citation & \cite{alexander_global_2018} \\
    
    \hline
    
    $\eta_{c,t}$ & NMR / (U5MR - NMR) \\
    
    \hline
    
    $g_1(\cdot)$ & $\log$ \\
    
    \hline
    
    Process model formula &
        $\log(\eta_{c,t}) = g_2(\cdot) + \epsilon_{c,t }$ \\
    
    \hline
    
     \multicolumn{2}{|l|}{\textbf{Covariate Component}} \\
    
    \hline
    
    $g_2(\cdot)$ & 
        $\beta_{c,0} + \beta_{1} \log(X_{c,t} - \log(\beta_2))\bm{1}_{[X_{c,t} > \beta_2]}$ \\
    
    \hline
    
    Covariates & 
        U5MR \\
    
    \hline
    
    \multicolumn{2}{|l|}{\textbf{Systematic Component}} \\
    
    \hline
    
    $g_3(\cdot)$ & 
        $\cdot$ \\
    
    \hline
    
    $\bm{\alpha}_c$ & 
        $\cdot$ \\
    
    \hline
     
    \multicolumn{2}{|l|}{\textbf{Offsets}} \\
    
    \hline
    
    $a_{c,t}$ &
        $\cdot$ \\
        
    \hline
    
    \multicolumn{2}{|l|}{\textbf{Smoothing Component}} \\
    
    \hline
    
    $\bm{B}$ & 
        $B_{c,t,k} = b_{c,k}(t)=$ cubic B-splines \\
    
    \hline
    
    $s(t_1, t_2)$ & 
        independent $k(t_1, t_2) = \sigma^2 1(t_1 = t_2)$ \\
    
    \hline
    
    $r$ & 1 \\
    
    \hline
    
    $\mathcal{K}_{d,c}$ & 
        $\left\{1, \dots, K_c \right\}$ \\
        
    \hline
    
    \multicolumn{2}{|l|}{\textbf{Parameter Estimation}} \\
    
    \hline
    
    Fixed & 
        $\cdot$ \\
    
    \hline
    
    Vague Priors & 
        regression coefficients \\
    
    \hline
    
    Informative Priors & 
        $\cdot$ \\
    
    \hline
    
    Hierarchical model & 
        smoothing parameters \\
     \hline
    
    Hierarchical distribution $\pi$ & 
        normal \\
    
    \hline
    
    Number of levels in hierarchy & 
        1 \\
    
    \hline
    
    Hierarchical groupings & 
        countries within world \\
        
    \hline
    
    \multicolumn{2}{|l|}{\textbf{Projections}} \\
    
    \hline
    
    Projections & 
    $\cdot$ \\
    
    \hline
\caption{TMMP specification for the Neonatal Mortality Rate model \citep{alexander_global_2018}.}
\label{tab-example-comparison}
\end{longtable}

\begin{longtable}[c]{
    |p{0.25\linewidth}
    |P{0.70\linewidth} |
} \hline
    & \textbf{Bmat} \\
    
    \hline
    
    Citation & \cite{alkema_bayesian_2017} \\
    
    \hline
    
    $\eta_{c,t}$ & proportion of non-AIDS deaths that are maternal among women of reproductive age \\
    
    \hline
    
    $g_1(\cdot)$ & $\log$ \\
    
    \hline
    
    Process model formula &
        $\log(\eta_{c,t}) = g_2(\cdot) + \epsilon_{c,t} $ \\
    
    \hline
    
     \multicolumn{2}{|l|}{\textbf{Covariate Component}} \\
    
    \hline
    
    $g_2(\cdot)$ & 
        $\beta_{c,0} + \sum_k X_{c,t,k} \beta_k$ \\
    
    \hline
    
    Covariates & 
        log(GDP), log(GFR), SAB \\
    
    \hline
    
    \multicolumn{2}{|l|}{\textbf{Systematic Component}} \\
    
    \hline
    
    $g_3(\cdot)$ & 
        $\cdot$ \\
    
    \hline
    
    $\bm{\alpha}_c$ & 
        $\cdot$ \\
    
    \hline
     
    \multicolumn{2}{|l|}{\textbf{Offsets}} \\
    
    \hline
    
    $a_{c,t}$ &
        $\cdot$ \\
        
    \hline
    
    \multicolumn{2}{|l|}{\textbf{Smoothing Component}} \\
    
    \hline
    
    $\bm{B}$ & 
        $B_{c,k} = b_{c,k}(t)=$ cubic B-splines \\
        $\bm{B} = \bm{I} $\\
    
    \hline
    
    $s(t_1, t_2)$ & 
        ARMA(1,1) \\
    
    \hline
    
    $r$ & 1 \\
    
    \hline
    
    $\mathcal{K}_{d,c}$ & 
        $\left\{ 1990 \right\}$ \\
        
    \hline
    
    \multicolumn{2}{|l|}{\textbf{Parameter Estimation}} \\
    
    \hline
    
    Fixed & 
        $\cdot$ \\
    
    \hline
    
    Vague Priors & 
        regression coefficients \\
    
    \hline
    
    Informative Priors & 
        $\cdot$ \\
    
    \hline
    
    Hierarchical model & 
        regression intercepts\newline
        smoothing parameters \\
        
     \hline
    
    Hierarchical distribution $\pi$ & 
        intercept: normal\newline
        smoothing: truncated normal \\
    
    \hline
    
    Number of levels in hierarchy & 
        intercept: 2 \newline 
        smoothing: 1 \\
    
    \hline
    
    Hierarchical groupings & 
        intercept: countries within region within world \newline
        smoothing: countries within world \\
        
    \hline
    
    \multicolumn{2}{|l|}{\textbf{Projections}} \\
    
    \hline
    
    Projections & 
    $\cdot$ \\
    
    \hline
\caption{TMMP specification for Bmat \citep{alkema_bayesian_2017}.}
\label{tab-example-comparison}
\end{longtable}

\begin{longtable}[c]{
    |p{0.25\linewidth}
    |P{0.70\linewidth} |
} \hline
    & \textbf{Mortality} \\
    
    \hline
    
    Citation & \cite{alexander_flexible_2017} \\
    
    \hline
    
    $\eta_{c,t}$ & age-specific mortality \\
    
    \hline
    
    $g_1(\cdot)$ & $\log$ \\
    
    \hline
    
    Process model formula &
        $\log(\eta_{c,t}) = g_2(\cdot) + \epsilon_{c,t}$ \\
    
    \hline
    
     \multicolumn{2}{|l|}{\textbf{Covariate Component}} \\
    
    \hline
    
    $g_2(\cdot)$ & 
         $\sum_k X_{a, k} \beta_{c, t, k} $ \\
    
    \hline
    
    Covariates & 
        $X_{k,a}$ is the $k$th principal component of the mortality schedule for age group $a$ \\
    
    \hline
    
    \multicolumn{2}{|l|}{\textbf{Systematic Component}} \\
    
    \hline
    
    $g_3(\cdot)$ & 
        $\cdot$ \\
    
    \hline
    
    $\bm{\alpha}_c$ & 
        $\cdot$ \\
    
    \hline
     
    \multicolumn{2}{|l|}{\textbf{Offsets}} \\
    
    \hline
    
    $a_{c,t}$ &
        $\cdot$ \\
        
    \hline
    
    \multicolumn{2}{|l|}{\textbf{Smoothing Component}} \\
    
    \hline
    
    $\bm{B}$ & 
        $\bm{B} = \bm{I} $\\
    
    \hline
    
    $s(t_1, t_2)$ & 
        independent \\
    
    \hline
    
    $r$ & 0 \\
    
    \hline
    
    $\mathcal{K}_{d,c}$ & 
        $\cdot$ \\ 
    \hline
    
    \multicolumn{2}{|l|}{\textbf{Parameter Estimation}} \\
    
    \hline
    
    Fixed & 
        $\cdot$ \\
    
    \hline
    
    Vague Priors & 
        $\cdot$ \\
    
    \hline
    
    Informative Priors & 
        $\cdot$ \\
    
    \hline
    
    Hierarchical model & 
        regression coefficients, smoothing parameters \\
        
     \hline
    
    Hierarchical distribution $\pi$ & 
        normal \\
    
    \hline
    
    Number of levels in hierarchy & 
        1 \\
    
    \hline
    
    Hierarchical groupings & 
        counties within state\\
        
    \hline
    
    \multicolumn{2}{|l|}{\textbf{Projections}} \\
    
    \hline
    
    Projections & 
    $\cdot$ \\
    
    \hline
\caption{TMMP specification for the age-specific mortality model \citep{alexander_flexible_2017}.}
\label{tab-example-comparison}
\end{longtable}

\bibliography{references}